\documentclass[a4paper]{cas-dc}

\usepackage[numbers,sort&compress]{natbib}
\usepackage{amsmath}

\def\tsc#1{\csdef{#1}{\textsc{\lowercase{#1}}\xspace}}
\tsc{WGM}
\tsc{QE}

\usepackage{lineno}

\begin{document}
\let\WriteBookmarks\relax
\def\floatpagepagefraction{1}
\def\textpagefraction{.001}

\shorttitle{Calibration and Performance of Germanium High Voltage Detectors for SuperCDMS SNOLAB}    

\shortauthors{SuperCDMS Collaboration, M.F. Albakry et.\ al}  

\title [mode = title]{Calibration and Performance of Germanium High Voltage Detectors for SuperCDMS SNOLAB}  



%


\author[1,2]{M.F.~Albakry}
\author[3]{I.~Alkhatib}
\author[4,5]{D.~Alonso-González}
\author[6]{J.~Anczarski}
\author[6]{T.~Aralis}
\author[7]{T.~Aramaki}
\author[2]{A.~Ashtari~Esfahani}
\author[8]{I.~Ataee~Langroudy}
\author[8]{R.~Bhattacharyya}
\author[8]{A.J.~Biffl}
% ...

\address[1]{Department of Physics \& Astronomy, University of British Columbia, Vancouver, BC V6T 1Z1, Canada} 
\address[2]{TRIUMF, Vancouver, BC V6T 2A3, Canada}
\address[3]{Department of Physics, University of Toronto, Toronto, ON M5S 1A7, Canada}
\address[4]{Departamento de F\'{\i}sica Te\'orica, Universidad Aut\'onoma de Madrid, 28049 Madrid, Spain}
\address[5]{Instituto de F\'{\i}sica Te\'orica UAM-CSIC, Campus de Cantoblanco, 28049 Madrid, Spain}
\address[6]{SLAC National Accelerator Laboratory/Kavli Institute for Particle Astrophysics and Cosmology, Menlo Park, CA 94025, USA}
\address[7]{Department of Physics, Northeastern University, 360 Huntington Avenue, Boston, MA 02115, USA}
\address[8]{Department of Physics and Astronomy, and the Mitchell Institute for Fundamental Physics and Astronomy, Texas A\&M University, College Station, TX 77843, USA}
% ...
















\begin{abstract}
As SuperCDMS SNOLAB is getting ready to search for low mass dark matter particles, using cryogenic Ge and Si detectors, a set of six of the new SuperCDMS High Voltage (HV) detectors (four Ge and two Si) were tested in the Cryogenic Underground TEst facility (CUTE) at SNOLAB. This provided the first opportunity to gain experience with this new detector type and assess their performance thoroughly under low background conditions. Here we describe the SuperCDMS HV detector concept and discuss some of the newly developed analysis methods and approaches. Focusing on the Ge detectors, we investigate the detector performance under voltage bias (up to 90~V), exercise the low energy (keV to sub-keV range) calibration based on the electron capture peaks generated by the decay of $^{71}$Ge, assess the detector resolution, and demonstrate the unexpected (and encouraging) ability of these detectors to also measure high energy interactions in the hundreds of keV range with good resolution (better than 3\% at 356~keV).
\end{abstract}




\begin{keywords}
 Dark matter \sep SuperCDMS \sep Cryogenic detectors \sep Calibration \sep CUTE \sep SNOLAB \sep NTL effect 
\end{keywords}

\maketitle

\section{Introduction}
\label{sec:intro}

SuperCDMS SNOLAB is the latest in a series of Cryogenic Dark Matter Search (CDMS) experiments (including CDMS I \cite{CDMSI}, CDMS II \cite{CDMSII} and SuperCDMS Soudan \cite{SuperCDMSSoudan}), using cryogenic semiconductor detectors to search for signals from dark matter particles. The success of the CDMSlite program \cite{CDMSlite1, CDMSlite2, CDMSlite3} that was executed as part of SuperCDMS Soudan prompted the development of a new type of detector, the SuperCDMS High Voltage, or HV, detector, which opens the door for the search for dark matter particles with masses well below 1~GeV/c$^2$. For a detector testing and characterization campaign, one set of six of these new detectors was operated at a temperature of $\sim$12 mK in the well-shielded Cryogenic Underground TEst facility (CUTE) \cite{CUTE_2024} at SNOLAB, a 2~km deep underground laboratory near Sudbury ON, Canada, from October 2023 until March 2024. 

The goal of this campaign was to provide a first opportunity to gain extensive experience with these new detectors, to assess their performance under low-background conditions and their response to operations under voltage bias, and to demonstrate energy calibration methods for electron interactions that can be implemented for SuperCDMS SNOLAB. This project was also used to exercise the full experimental chain from detector to recorded data; to test the new SuperCDMS data acquisition system (DAQ), the data transfer, and the processing scheme; and to develop data processing and analysis tools and methods for this new detector design. 

This paper focuses on the calibration and performance of the Ge detectors that were part of the detector testing campaign in CUTE. In Section~\ref{sec:hv_detectors}, the HV detector design and the detector readout scheme are discussed. Section~\ref{sec:calib_methods} describes the calibration methods. The details of data collection and processing and the basic analysis steps are laid out in Section~\ref{sec:data_acq_process}. The calibration and detector performance assessment are described in Section~\ref{sec:calib}.

\section{HV Detectors and Their Readout}
\label{sec:hv_detectors}
A sufficiently energetic particle interaction in a semiconductor produces lattice vibrations (phonons) and promotes electrons from the valence to the conduction band, generating electron-hole (or eh) pairs. Generally, it is assumed that when the charge carriers recombine, the energy expended to generate them is converted to phonons. If this is the case, then the phonon signal should reflect the total interaction energy.

Traditionally, CDMS and SuperCDMS detectors were based on Ge or Si crystals equipped with sensors to measure both phonon and charge signals. The ratio of the two signals can be used to distinguish between nuclear and electron interactions, based on different efficiencies for producing eh pairs. This enabled almost background-free searches for dark matter particles interacting via nuclear recoils with energies of more than about 10~keV \cite{CDMSI,CDMSII,SuperCDMS_HT}. However, relaxing the goal of a background-free measurement made it possible to noticeably reduce the energy threshold to include interactions with energies as low as 2~keV \cite{SuperCDMS_LT}.

During SuperCDMS Soudan, a new mode of operating the SuperCDMS detectors was tested and implemented for individual detectors whereby a high voltage bias of several 10s of volts (to be compared to standard operations with only a few V) was applied across the detector substrate \cite{CDMSlite1,CDMSlite2,CDMSlite_R2long,CDMSlite3}. As the charge carriers drift toward the electrodes, they quickly reach their terminal drift velocity and convert the electric potential energy into phonons. This process is referred to as the Neganov-Trofimov-Luke (NTL) amplification \cite{Neg_Trof,Luke}, and the phonons produced by the drifting charges are called NTL phonons. The total phonon energy $\rm E_t$ is then given by
\begin{equation}
    \label{eq:NTL-gain}
    {\rm E_t} = {\rm E_i} + ne\rm V_b,
\end{equation}
where $\rm E_i$ is the energy transfer in the primary interaction, $n$ is the number of eh pairs produced in the interaction, $e$ is the electric charge of the electron, and $\rm V_b$ is the applied bias voltage. The number of eh pairs depends on the average pair-creation energy $\epsilon$, which is 3~eV for electron interactions in Ge \cite{Pehl_1968_epsilon}. Therefore, we can rewrite Eq.~\eqref{eq:NTL-gain} as
\begin{equation}
    \label{eq:NTL-amp}
    {\rm E_t} = {\rm E_i} \left(1 + \frac{e\rm V_b}{\epsilon}\right).
\end{equation}

To account for the smaller number of eh pairs produced by nuclear interactions when compared to electron interactions, it is common to add an energy dependent yield factor in front of the fraction inside the parentheses in Eq.~\eqref{eq:NTL-amp} (see e.g.\ Eq.\ (7) in \cite{CDMSlite_R2long}) instead of modifying $\epsilon$. For the purpose of this publication, however, we are only concerned with electron interactions, and hence we have omitted this factor.

The internal amplification mechanism described by Eq. \eqref{eq:NTL-amp} increases the total phonon energy for a given interaction energy. Because the detection threshold is a threshold for the total phonon energy, this lowers the threshold for the interaction energy. However, for voltages that are high compared to $\epsilon/e$ (high voltage operation), the total phonon energy is dominated by the NTL phonons, which turns the phonon measurement \textit{de facto} into a charge measurement. This in turn means that the power to discriminate between electronic and nuclear interactions is lost.

For SuperCDMS SNOLAB, a new set of detectors was designed and fabricated with this HV mode of operation in mind. These HV detectors are optimized for energy resolution, and their electric field configuration is more uniform than that of SuperCDMS detectors with both phonon and charge readout. The main design considerations and features of the HV detectors are summarized below. A more detailed discussion can be found in references \cite{Kurinsky:20173w} and \cite{kurinskyThesis}.

\begin{figure}[pos=h]
    \centering
    \includegraphics[width=0.95\linewidth]{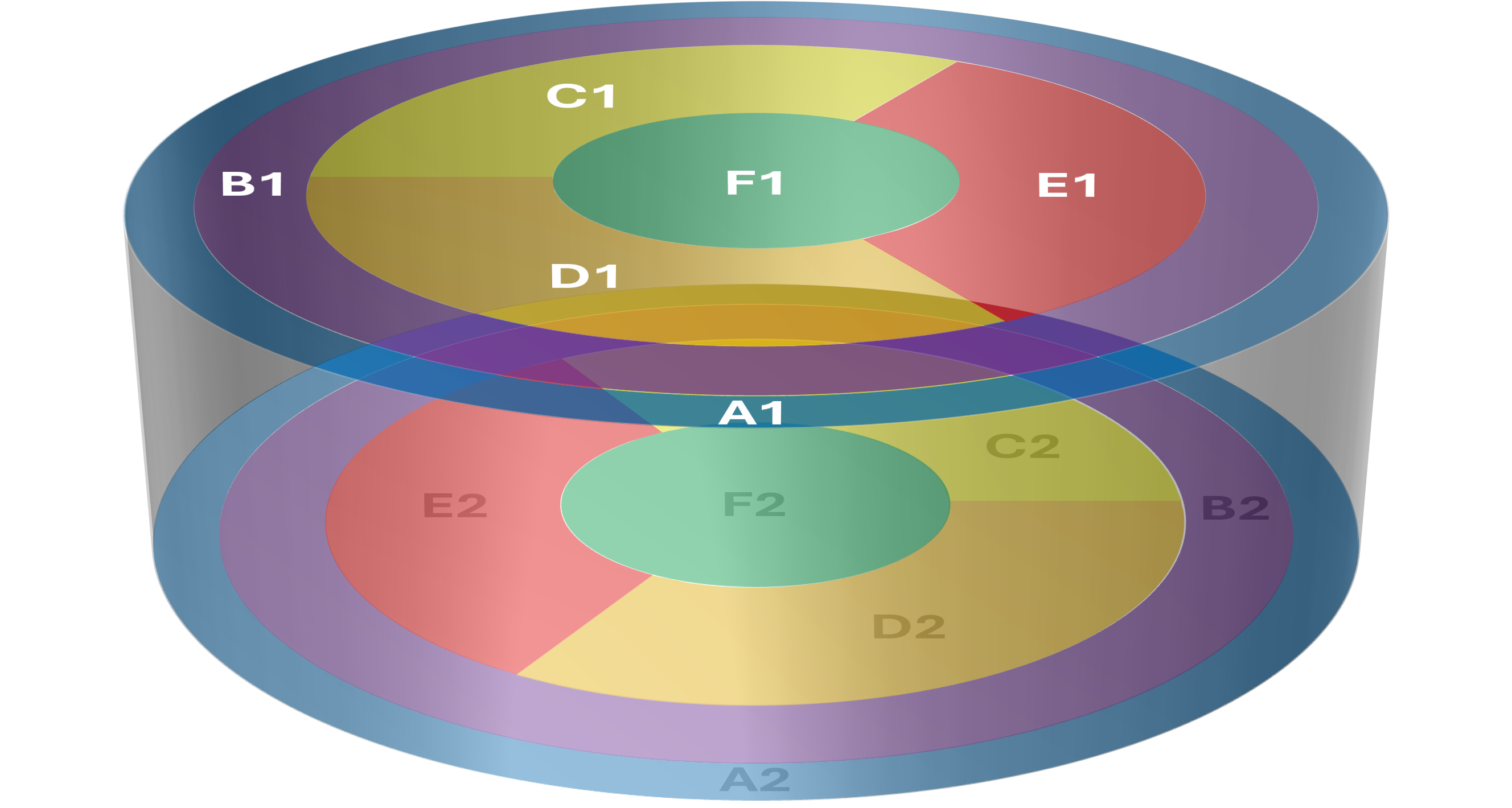}
    \caption{Channel layout of the SuperCDMS HV detectors. Each colored section represents one phonon channel: on each side, there are two outer ring channels, A and B, three wedge channels, C, D and E, arranged counter-clockwise (when viewed from above the face), and one circular channel, F, in the center (the channel and side labels are shown in the sketch).}
    \label{fig:HV_channels}
\end{figure}

\begin{figure}[pos=h]
    \centering
    \includegraphics[width=0.9\linewidth]{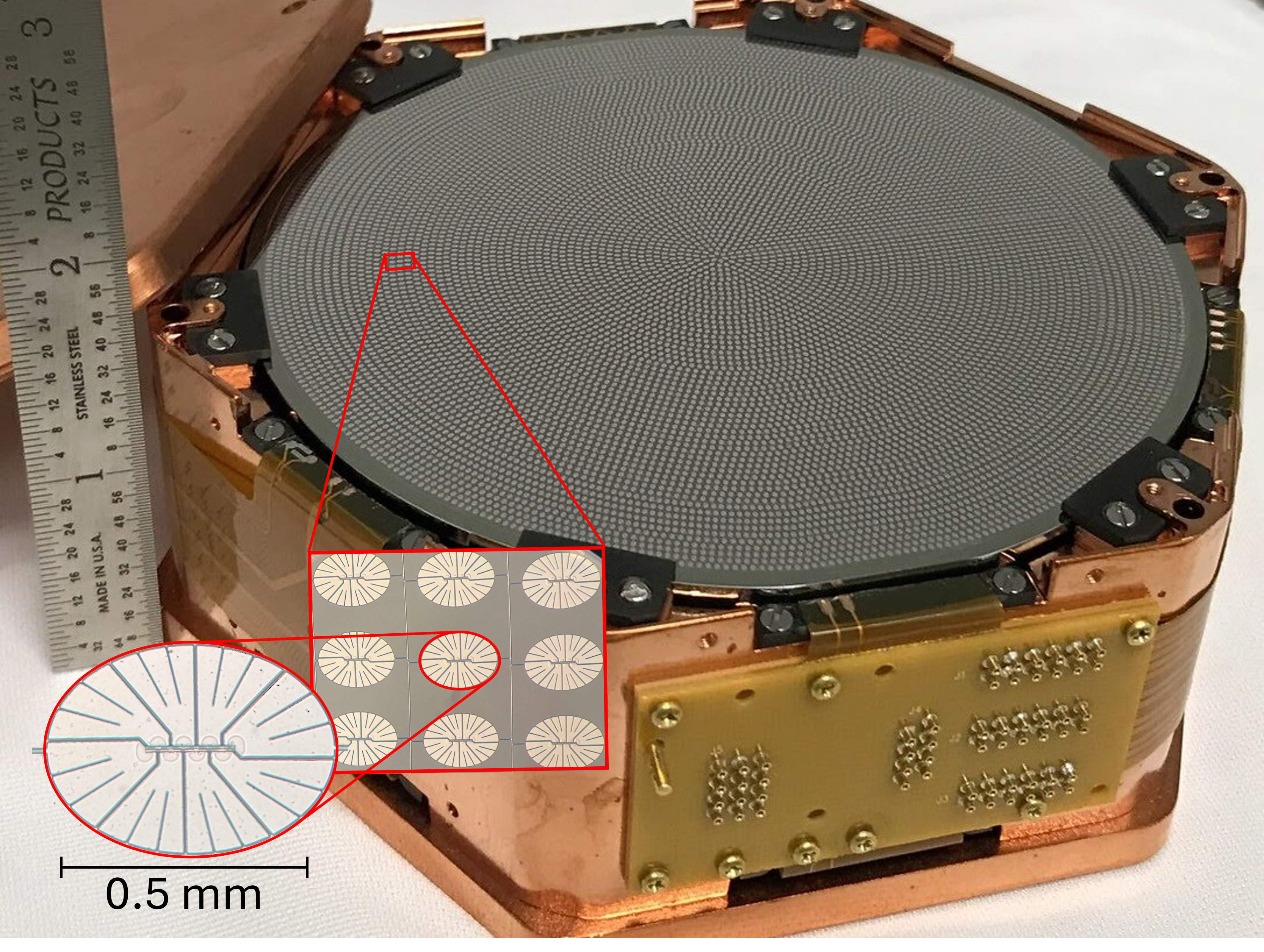}
    \caption{HV detector in its housing. The first zoom shows a section of nine QET cells. The second zoom shows an individual cell with the tungsten TES surrounded by the aluminum fins.}
    \label{fig:HV_pic}
\end{figure}

The SuperCDMS SNOLAB detectors are based on cylindrical Ge or Si crystals with a diameter of 100~mm and a height of 33~mm and thus have more than twice the volume of the SuperCDMS Soudan detectors (75~mm in diameter and 25~mm in height). The phonon detection is based on the traditional CDMS and SuperCDMS concept of the Quasiparticle-trap-assisted Electrothermal feedback Transition edge sensor (QET, see e.g.~\cite{QET_Irwin95, NIMA_444_2000_300}). Small patches of superconducting aluminum (called fins) collect phonons from the crystal bulk, which results in the breaking of Cooper pairs. The unpaired electrons (usually referred to as quasiparticles) diffuse through the aluminum, and when they reach the interface to the tungsten transition edge sensor (TES) where the superconducting band gap is reduced, they are trapped and eventually deposit their remaining excess energy into the TES. A bias current across a small (5~m$\Omega$) shunt resistor sets an approximately constant bias voltage for the TES which is adjusted so that the current through the TES heats it to its operating point about halfway between the superconducting and normal states. Energy deposited into the TES temporarily increases its resistances, leading to a reduction in TES current and hence the heating (electrothermal feedback). This is the main mechanism for removing excess energy from the system as the TES recovers to its baseline operating point. The change in current is sensed by an amplifier circuit based on Superconducting Quantum Interference Devices (SQUIDs) and forms the signal that indicates an interaction in the detector. The Al fins are more than 300 times larger than the TES itself. This results in a fast collection of phonons without having to increase the size (and thus the heat capacity) of the TES. The QETs are distributed over both flat faces of the detector and grouped into six readout channels on each face (see Fig.~\ref{fig:HV_channels}), with 1833 QETs per channel connected in parallel. Figure~\ref{fig:HV_pic} shows a picture of a SuperCDMS SNOLAB HV detector with an inset that zooms in on a small number of QETs and then a single QET cell. The sensor layout is optimized for fast and efficient phonon collection while minimizing the TES volume, and thus the sensor heat capacity.

The detectors are arranged in so-called towers. A tower consists of a stack of six detectors attached to a structure (tower body) that provides mechanical support, as well as thermal and electrical connections, and the central elements of the first-stage amplifiers. One of these towers (referred to as Tower 3), composed of four Ge and two Si HV detectors, was operated in CUTE during the campaign discussed here.

The readout scheme used in the CUTE campaign is identical to the one designed for SuperCDMS: the readout cables inside the cryostat (one per detector) terminate at a custom-designed printed circuit board, the Vacuum Interface Board (VIB), which is part of the vacuum seal and provides filtered electrical connection between the inside and the outside of the cryostat. Each HV detector is operated via two Detector Control and Readout Cards (DCRCs, one for each side of the detector) that are connected directly to the VIB. With the DCRC ground plane not connected to any external ground reference, it is possible to apply a voltage of up to $\pm$50~V to each DCRC (limited by the filter components on the VIB), allowing for a voltage bias across the detector of up to 100~V. The DCRCs include an analog section for sensor operation and signal amplification, a signal digitizer section and a transformer section; Ethernet is used for all communications (including synchronization between different DCRCs), data transfer, and for providing power from a Power-over-Ethernet (POE) unit. The voltage bias is provided by an external high voltage power supply via a connection at the VIB. The continuous data stream is initially stored locally in a large circular buffer. Events are identified using a configurable trigger logic programmed into the FPGA on the DCRC. Our default trigger algorithm is based on an Optimal Filter (OF) \cite{OF_description}, using signal template and noise information provided by the user \cite{SCMDS_L1trigger}. When a trigger is issued, the data for the corresponding event (traces) are written to disk. In addition to the triggers arising from the detector signals, we implement random triggers to collect signal-free data that are used
to assess the noise in the measurement. A description of an early prototype of these readout cards can be found in Ref.~\cite{Hansen_2010_DCRC}.

For this campaign, data were recorded with a sampling rate of 625~kHz. An event is 52~ms long and consists of a $\sim$26~ms long pre-trigger region (also referred to as pre-pulse baseline) and an equally long signal or pulse region. With typical signal fall times no longer than a few ms, this ensures that pulses (except those with very high energy) are fully contained within the recorded trace.

\section{Energy Calibration Methods}
\label{sec:calib_methods}

A low energy calibration was performed using the e\-lec\-tron capture (EC) decay of $^{71}$Ge which has a half-life of 11.5~days \cite{Ge-71_ref1, Ge-71_ref2, Cai_Ge71}. This isotope is produced \textit{in situ} through neutron activation, primarily of $^{70}$Ge, using neutrons emitted by a $^{252}$Cf source. The EC decay generates characteristic emission lines corresponding to the binding energy of the captured electrons in the Ga daughter. The most prominent emissions follow the K-shell, L1-shell and M1-shell captures \cite{x-rays_Ge71_1, x-rays_Ge71_2} at probabilities of roughly 88\%, 10\% and 1.5\% \cite{Manduchi_Ge71_EC_ratios}, producing spectral lines that correspond to energies of 10.37~keV, 1.30~keV and 160~eV, respectively, which are used to determine the conversion of signal amplitudes to interaction energies.

A high energy calibration was performed using a $^{133}$Ba source, which produces a wide range of gamma lines, the most prominent of which are at 356.0~keV, 302.9~keV, 383.9~keV and 276.4~keV \cite{ba-gamma1, ba-gamma2}. For Ge detectors of the given size, the probability of full absorption of these gammas is high enough to generate observable full energy absorption peaks in the spectrum that can be used for calibration.

\section{Data Collection and Processing}
\label{sec:data_acq_process}

\subsection{Measurement Periods and Detector Conditioning}
\label{sec:data}
\vspace{0.5em}
The Tower 3 testing campaign included four separate cooldown cycles (referred to as Runs 35 through 38). Run 35 only lasted a few days due to a thermal short between the 4K and 60K stages of the cryostat, making detector operation difficult. Run 36 started after fixing the thermal short. It lasted for about a month and was terminated due to anticipated service outages at SNOLAB. However, the payload only warmed up to $\sim$30~K for about a day before cooling back down for Run 37. This run ended with a warm-up to room temperature after about three months due to operational issues with the dilution refrigerator. Run 38 lasted about one month, until the end of the testing campaign.

The measurement time was divided between general functionality tests of all six detectors in the payload, dedicated campaigns to understand and minimize sources of excess noise, measurements with an external $^{133}$Ba source (for Si detector calibration and the high energy calibration of the Ge detectors) and measurements without the $^{133}$Ba source (referred to as low-background measurements; for the assessment of detector and facility background, as well as Ge low energy calibration). Both, low-background and Ba measurements, were taken with and without voltage bias, referred to as HV data and 0~V data, respectively.

After an initial cooldown, the charge transport in the detectors is severely impeded by an abundance of charged impurities. These need to be neutralized, which in SuperCDMS is accomplished by illuminating the detectors with near-infrared (IR) photons ($\sim$940~nm) emitted by LEDs mounted inside the detector housing. After a series of initial tests in Run 36, mostly without voltage bias, a first neutralization campaign was launched at the beginning of Run 37. 

The extended operation of the detectors with voltage bias can lead to accumulation of space charge that may reduce NTL amplification or cause localized breakdown near the electrodes. The space charge can be cleared by short, relatively low-intensity bursts (flashes) of photons from the IR-LEDs. After the initial neutralization campaign, the detectors were flashed regularly (typically every 2-3~days) to ensure consistent detector performance.

As previously observed during the CDMSlite campaigns of SuperCDMS Soudan \cite{CDMSlite1,CDMSlite2,CDMSlite3}, applying a high bias voltage to neutralized detectors leads to an initial high rate of low energy events, consistent with weakly bound charge carriers being pulled away from their bound states by the electric field via a tunneling process (internal leakage). At higher voltages, the total energy (internal leakage rate times NTL gain) is often so high that the sensors go fully through the transition into their normal conducting state. The time constant for a full recovery can be tens of minutes. However, if, after waiting for about 5-10 minutes, the voltage is dropped slightly (by a few volts), the event rate and operating point typically return instantly to their nominal values. Therefore, all data with voltage bias were taken after an initial pre-biasing period of about 10~minutes during which the voltage bias had been set to between 5~V and 10~V above the measurement voltage.

The four Ge detectors in the tower are in positions 1, 3, 4 and 6 (where 1 is closest to the tower body and 6 is farthest away). Detectors 1 and 6 were fully functional; for detector 3, it was not possible to find a viable operating point for channel A1, the outermost (ring-)channel on side 1; hence, for this detector all data were taken with only the remaining eleven channels. Detector 4 exhibited multiple problems with cabling and pre-amplifiers, so we decided to not take data with it. It is conceivable that some or all of these issues can be resolved for SuperCDMS SNOLAB. 

All calibration data reported here were taken after neutralization, during Runs 37 and 38. During Run 37, we took extended data sets at 0~V (all three working detectors) and 50~V bias (only detectors 1 and 3), both for low energy and high energy calibration. During Run 38, we took a range of shorter data sets at bias voltages ranging from 0~V to 90~V (all three detectors). Data for assessing the noise were taken throughout the measurement campaign. 

\subsection{Template Generation}
\label{sec:templates}
\vspace{0.5em}
Our main event reconstruction algorithm is an OF \cite{OF_description, sig-proc-SSdet}. This algorithm fits a signal template, representing the pulse shape, to the raw trace in the frequency domain, while accounting for the relative power of signal template and noise in each frequency bin. 

The template required by the OF is constructed from an average of good pulses after aligning the start times. Pulses where at least one of the channels absorbs enough energy so the TES is driven near or into its normal conducting state, leading to a significant distortion of the shape (called saturated pulses) are excluded from the averaging, as are low energy pulses with amplitudes near the noise floor. Templates are constructed for each channel individually, as well as for a sum trace, which is constructed by summing all individual channel traces after accounting for the channels' relative sensitivities (see Sec.~\ref{sec:rel_cal}). We initially process data using a template with an approximate shape given by an exponential model with one rise and one fall time, with estimated model parameters. This allows us to select an appropriate set of traces for construction of the final template. The pulses from the different events are averaged after aligning them in time. For individual channel traces, the alignment uses the start time of the pulse, calculated by the initial processing. The sum traces are aligned instead on the rising edge at 30\% of the pulse height. The pulse average is then fitted by an empirical function of time, $f(\text{t})$, with four exponential terms, representing one rise and three fall times:
\begin{equation}
        f(\text{t}) = a_1e^{-\frac{(\text{t}-\text{p})}{\text{t}_1}} + a_2e^{-\frac{(\text{t}-\text{p})}{\text{t}_2}} + a_3e^{-\frac{(\text{t}-\text{p})}{\text{t}_3}} + a_4e^{-\frac{(\text{t}-\text{p})}{\text{t}_4}}, 
    \label{eq:template_fit}
\end{equation}
where $a_1 = -(a_2+a_3+a_4)$, and $a_2$, $a_3$ and $a_4$ are the amplitudes corresponding to the three fall times, p is the start time of the pulse, $\text{t}_1$ is the rise time, and $\text{t}_2$, $\text{t}_3$ and $\text{t}_4$ are the three fall times. The fits are renormalized to ensure an amplitude (pulse maximum) of 1, and shifted so that the start time of all template pulses is at 26~ms to align the pulse start time with the trigger point used in the DAQ system. Template values prior to this point are set to 0. Figure~\ref{fig:template} shows an example template (detector 1, 50 V data set) together with individual pulses and the pulse average that was used to generate the template.

\begin{figure}[pos=h]
    \centering
    \includegraphics[width=0.9\linewidth]{figs/Template_det1_50V.png}
    \caption{Template for detector 1 for the 50 V data set (red); also shown are 30 individual low-pass filtered pulses (gray) randomly selected from the set of pulses used for generating the template and the average (blue) of all 333 pulses that were used for the template generation. The template given by the functional form fit (see text) is a very good representation of the pulse average.}
    \label{fig:template}
\end{figure}

\subsection{Event Reconstruction}
\label{sec:reconstruction}
\vspace{0.5em}
The energy deposited during a given event is estimated based on the OF amplitude of the sum trace. The individual channels' amplitudes and start times carry information about the location of the interaction inside the detector.

Traces recorded in the absence of an interaction in the detector (in response to the random triggers mentioned in Sec.~\ref{sec:hv_detectors}) are used to measure the baseline resolution, which corresponds to the uncertainty in determining the pulse amplitude. When searching for a pulse in the data, the OF algorithm picks the start time that returns the largest amplitude, which corresponds to the best fit. However, in the absence of a pulse, this biases the reconstructed amplitude values away from zero. For a fair assessment of the baseline resolution, we therefore employ a modified version of the OF algorithm where the start time is fixed to the nominal trigger point (zero-delay OF). We then define the baseline resolution as the standard deviation of the Gaussian function that best fits the resulting amplitude distribution. 

At high energies, typically somewhere above 100~keV of total phonon energy, sensor saturation leads to a distortion of the pulse shape. In this regime, the OF amplitude does not scale linearly with the energy absorbed by the sensor anymore, and above some energy (typically several hundred keV) it becomes fully degenerate. Therefore, at high energy, we use the integral of the sum trace as a basis for our energy estimation.

\subsection{Relative Calibration}
\label{sec:rel_cal}
\vspace{0.5em}
The detector response varies across channels, resulting in different
amplitudes for the same amount of absorbed energy. This is due to differences, for example, in TES characteristics or parasitic power between channels. To account for this, each channel is assigned a weight or relative calibration factor. We first determine the relative calibration factors for the sensors on a given side of the detector and then determine the relative weight between the two sides. 

To accomplish this, we project the non-saturated events from a Ba calibration data set with amplitudes well above the baseline noise onto a plane using the pulse amplitude information from three channels at a time. The corresponding $x$ and $y$ coordinates for channels ($ijk$) on side 1 of the detector are defined as follows:

\begin{equation} \label{eq:rel_calib_xy}
\begin{split}
    x_{ijk} = \frac{w_i A_i \cos(0^\circ) + w_j A_j \cos(120^\circ) + w_k A_k \cos(240^\circ)}{w_i A_i + w_j A_j + w_k A_k} 
\\
y_{ijk} = \frac{w_iA_i\sin(0^\circ)+w_j A_j \sin(120^\circ) + w_k A_k \sin(240^\circ)}{w_i A_i + w_j A_j + w_k A_k}
\end{split}
\end{equation}

Here, $A_i$ is the observed OF amplitude in channel $i$, and $w_i$ is the corresponding weight or relative calibration factor. For side 2, the sign of the $x$-coordinate is flipped but the $y$-coordinate is identical. The relative calibration factors are then adjusted until the relevant features of the distribution match the expectation.

We start with the wedge channels (C, D, E) on one side of the detector. Given the three-fold symmetry of the channel layout (see Fig.~\ref{fig:HV_channels}), we expect to find this symmetry also in the outline of the event distribution based on the above projection. If events in the medium energy range (well above noise and away from saturation) are chosen, the outline of the observed distribution is a triangle; the weights are then adjusted so that the triangle becomes equilateral and the corners align with the main axes of the above projection. 

If we replace one of the three channels in the above triangle plot by channel A, B or F, the symmetry of the channel layout is broken, and the distribution of events is now a distorted and somewhat displaced triangle. In this case, it is not obvious how the distribution should be aligned. Here, we take advantage of a Detector Monte Carlo (DMC) simulation package based on G4CMP \cite{G4CMP3}, a publicly available \textsc{Geant4} \cite{Geant, G4CMP1, G4CMP2} module that was originally developed by the SuperCDMS collaboration. The DMC simulations track charge carriers and phonons throughout the detector and model their absorption in the sensors, and the sensor response, producing output files that have the same format and are processed in the same way as data coming from the detectors. The weight of the new channel (A, B or F) is then adjusted until the outline of the measured distribution best matches the DMC-generated distribution, which is based on monoenergetic events uniformly distributed throughout the detector.

\begin{figure}[pos=h]
    \centering
    \includegraphics[width=0.90\linewidth]{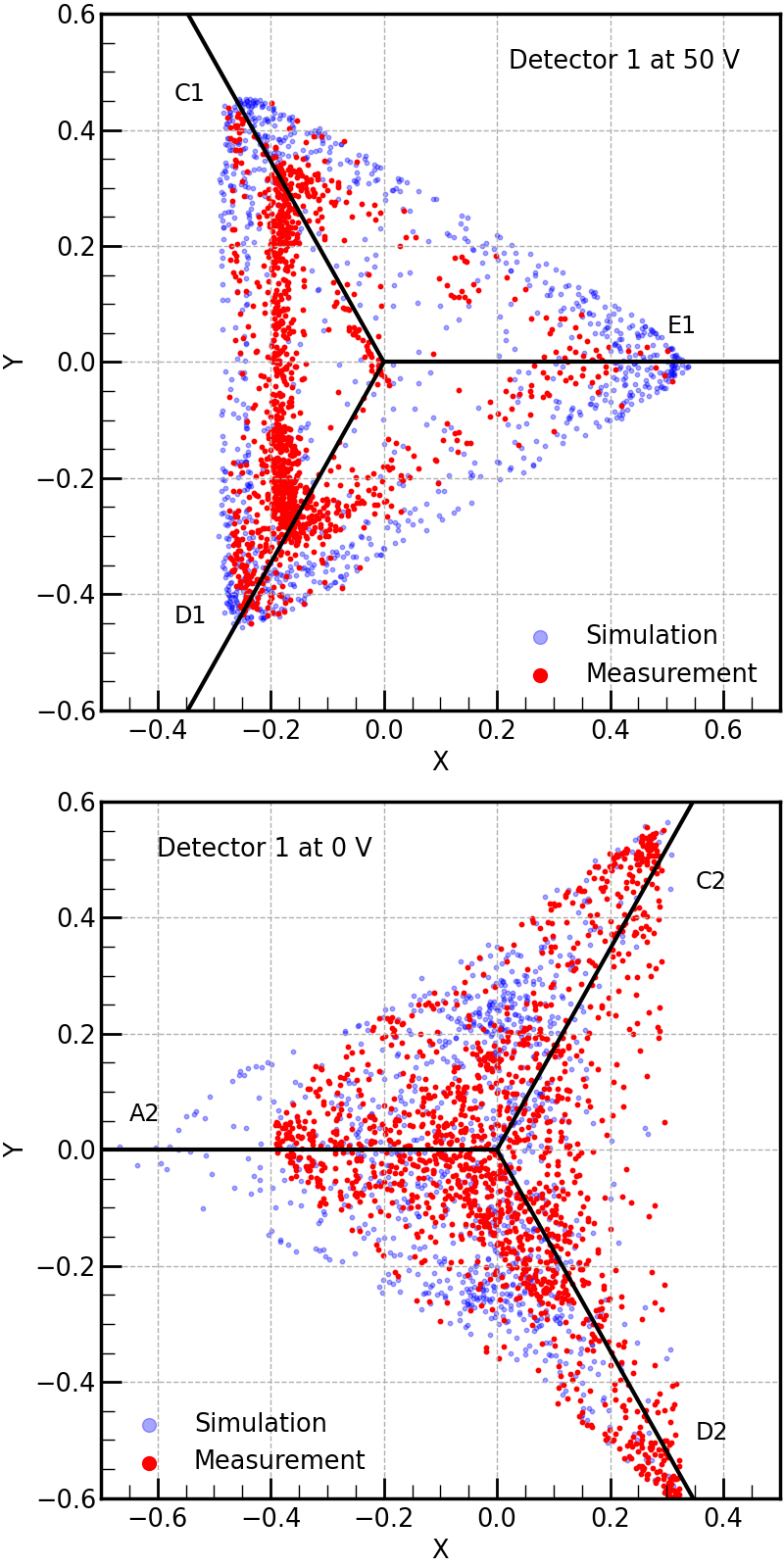}
    \caption{Comparison of simulated (blue) and measured (red) event distributions according to Eq.~\eqref{eq:rel_calib_xy}, after alignment. Measurements were taken with detector~1 exposed to a $^{133}$Ba source. The black lines indicate the symmetry imposed by the construction of the coordinates. {\it Top:} Three wedge channels (CDE) on side~1 at 50~V. Despite the inhomogeneous event distribution, the outline of the (equilateral) triangle in the measured data is clear. {\it Bottom:} Side~2 at 0~V including the outermost channel (CDA). Measurement and simulation differ near the tip of the triangle at A. However, the outlines of the distributions above an x-value of about -0.2 line up well, indicating that the simulation fairly represents the measured data in this region and hence that this method is still applicable despite shortcomings of the simulation in some specific aspects.
    }
    \label{fig:triangle_plots}
\end{figure}

Figure~\ref{fig:triangle_plots} shows examples of event distributions for three wedge channels (CDE) and a ring channel with two wedge channels (CDA) together with simulated distributions\footnote{Simulations were performed using \textsc{Geant4} version 10.7.4 and G4CMP V09-00-00}.

The simulations are still under active development and not fully validated yet. It is therefore no surprise that there are some differences between DMC results and measured data. Perhaps the most striking difference is that DMC predicts a stronger response for the outer channels for events occurring near the edge of the detector. However, since the response of the outer channels to events closer to the center of the detector appears to be compatible between DMC and data, the method is still applicable if the respective features of the distributions are aligned. In the example in Fig.~\ref{fig:triangle_plots}, bottom, the tip of the triangle that corresponds to channel A is under-populated in measured data when compared to the DMC predictions, but the position and shape of the triangle base (positive x-values), which is populated by events closer to the center of the detector, are the same between measured data and DMC, and that is what is used to determine the relative calibration factor here. By following the same approach for distributions that include either channel B or channel F, we can determine the relative calibration factors for all channels on a given side.

After both sides are balanced individually, we balance the two sides against each other by looking at similar projections where we use wedge channels from both sides (C and D from one side and E from the other side) and then scale all relative calibration factors on a given side by the same factor to reach overall balance between the two sides while maintaining the balance on each side individually.

Generally, the distributions have more pronounced substructures in 50~V data than in 0~V data. This is due to the more localized energy deposition from the NTL phonons, which are produced in columns throughout the detector, and as such on average closer to the phonon sensors than the phonons from the primary interaction site. These differences are well reflected in the corresponding DMC data sets. This enhances the confidence that this method leads to a reliable relative calibration between the channels, even though the DMC does not perfectly reproduce the measured data. Imperfections in the method may lead to a worsening in the resolution and a slightly larger than necessary residual position dependence of the energy response, but do not introduce a systematic bias on the peak positions.

The strength of this method lies in its universal applicability as it does not require mono-energetic event samples or events that are homogeneously distributed within the detector. As long as there are enough events that are well above the detection threshold and below the saturation range in all channels and that are spread across some reasonable fraction of the detector, enough of the shape will be mapped out so that the relative calibration factors can be determined.

\subsection{Event Selection}
\label{sec:cuts}
\vspace{0.5em}
We apply a series of event selection criteria to ensure that only good quality data are included in our final event sample. These criteria can broadly be grouped into three categories: basic criteria, which ensure that we deal with events of the type we intend to study; quality criteria, which remove events where the raw traces are not compatible with the features expected from a single particle interaction, or that occurred during a time when the detector was not operating properly; and focus criteria which select a subset of events that are of interest for the specific analysis in question. Below, we give a short description of the different criteria in the three categories.

\vspace{0.5 em} Basic selection criteria include:
\begin{itemize}
    \item {\it Trigger type}: For baseline resolution analyses, we select events recorded due to random triggers generated by the DAQ, while for all other analyses, we only select events that were triggered by our particle interaction trigger algorithm.
    \item {\it Triggering detector}: Here, we ensure that only those data are included that are triggered by the detector under consideration.
    \item {\it No empty traces}: For synchronization purposes, our DAQ regularly produces event entries in the data stream that do not contain any trace information. This criterion removes all events without trace information.
    \item {\it No duplicate events}: We discovered during the measurement campaign that due to a problem with the trigger system (which has since been resolved), in some cases the DAQ records the same event twice. We remove the additional copy of these duplicate events.
    \item {\it No LED events}: We remove data periods that coincide with the internal LEDs being operated.
\end{itemize}

Data quality selection criteria include:
\begin{itemize}
    \item {\it Good data period}: This criterion removes -- for each detector separately -- periods of time where the detector shows a clear deviation from regular behavior such as a sudden change in noise conditions or an individual channel railing (i.e.,\ the output value is constantly at the amplifier limit), which can occur e.g.\ when the sensor current changes faster than the SQUID amplifier circuit can follow.
    \item {\it Trigger time}: If configured correctly, all events should trigger roughly at the same time within the recorded trace. This nominal trigger point was chosen to be in the middle of the recorded traces at $\sim$26~ms. However, occasionally, we observe shifted events appearing several ms early or late in the trace. This is attributed to an artifact of the OF triggering algorithm: the OF-filtered trace often shows satellite peaks at specific intervals before and after the main peak; for large pulses, these satellite peaks can exceed the trigger threshold. To ensure that all those events are properly captured, we allowed the event reconstruction algorithm to search for the pulse up to 20~ms away from the nominal trigger point. However, since the OF algorithm returns the largest amplitude within the search window, it can happen that a low amplitude event generates a trigger and appears at the nominal trigger point, but in individual channels there are noise excursions elsewhere in the trace with a larger amplitude, so that the OF returns these noise fluctuations rather than the event that triggered. Therefore, we require a small delay at low energy and widen the window at high energy to include the shifted events. The width of the window at high energy may differ from data set to data set since the time shift of the off-set triggers varies.
    \item {\it Good baseline}: We remove events with abnormally large fluctuations in the baseline before or after the pulse, or with a clear slope in the baseline. 
    \item {\it No pileup}: We aim to reject events with multiple pulses within one event trace. To this end, a peak search algorithm is applied to each channel's trace individually, and an event is removed if the average number of peaks found per channel is above 1.5.
    \item {\it Low $\chi^2$}: When a pulse is fitted by the OF algorithm, the quality of the fit is determined by calculating the squared difference between template and pulse ($\chi^2$) in the frequency domain. However, only frequencies below 40~kHz are considered (low frequency or LF-$\chi^2$). A large majority of events populate a well defined LF-$\chi^2$ distribution. Outliers are removed. 
    \item {\it Non-negative amplitudes}: In some data sets, a population of events appears in detector 3 that do not have the typical pulse shapes and show large negative amplitudes in some channels (though the frequency content is sufficiently pulse-like so they pass the $\chi^2$-test). These events are removed by requiring that none of the channels have a large negative amplitude.
\end{itemize}

Finally, when a voltage bias is applied, the electric field near the edge of the detector is distorted so that charge carriers in some regions do not move through the full potential and therefore show reduced NTL amplification. Hence, for data sets taken under non-zero voltage bias, we define a focus criterion that excludes events near the edge of the detector:
\begin{itemize}
    \item {\it Low radius events}: The true radial position of an event is not known, but the layout of the sensors together with the mechanisms of phonon propagation and absorption allow us to define a parameter \mbox{$\text{r}_\text{p}$ = (PA -- PF)/PT} that is related to the radial position (radial parameter), where PA and PF are the sum of the OF amplitudes in the outer and inner channels on both sides of the detector, respectively, and PT is the amplitude of the sum trace. The low radius event selection is then given by the condition r$_\text{p} <$ 0.025. This threshold was chosen to reject most, if not all, events with reduced NTL amplification while maintaining sufficient statistics in the activation peaks. For detector 3 where the outer channel on side 1 is not working, the numerator in the definition of r$_{\rm p}$ uses only the information from side 2, and the selection condition for low radius events is therefore replaced by r$_\text{p} <$ 0.025/2.
\end{itemize}

\section{Detector Calibration}\label{sec:calib}
\subsection{Low Energy Calibration}\label{sec:low_calib}
\vspace{0.5em}
For the low energy calibration, we use data taken in the days following the irradiation of the payload with neutrons from a $^{252}$Cf source. After applying all basic and data quality selection criteria, and for HV data also the low radius event selection, we can identify peaks in the energy spectrum that originate from the decay of $^{71}$Ge. 

At 0~V, the K-shell and L-shell capture events at 10.37~keV and 1.30~keV, respectively, are clearly visible as peaks in the energy spectra of detectors 1 and 3. For detector 6, only the K-shell peak is visible; the L-shell peak hides in the noise distribution. At 50~V bias (only detectors 1 and 3), the M-shell peak at 160~eV is also visible and is included in the calibration.

To determine the peak positions, we fit a Gaussian function to the data, using an unbinned negative log-likelihood (NLL) approach. This requires that the contribution of non-peak events in the fit range be negligible. This is easily accomplished in most cases given the overall low background rate. Only for detector 3 at 0 V a window size of less than 2.5~$\sigma$ had to be chosen to avoid contamination of the fit region by events that had a significant probability of not originating from $^{71}$Ge.

\begin{figure}[pos=h]
    \centering
    \includegraphics[trim={2cm 1cm 3.2cm 3.5cm},clip,width=0.98\linewidth]{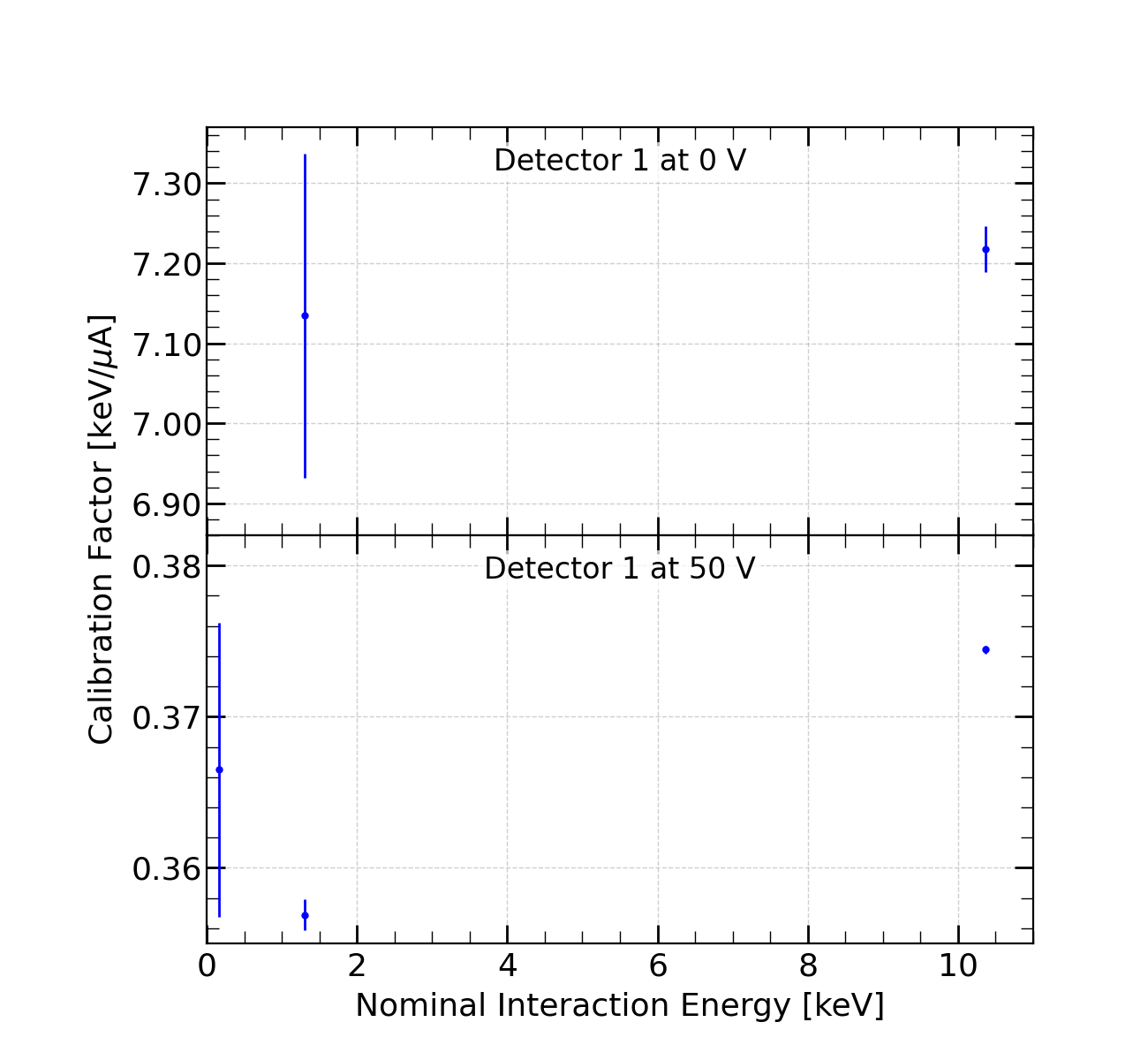}
    \caption{Calibration factors deduced from the positions of the individual $^{71}$Ge EC peaks in the amplitude spectrum of detector 1 for both the 0~V and the 50~V data. At 0~V, the calibration factors deduced from the L- and K-shell positions are consistent with each other, which means that the peak positions are consistent with a linear energy calibration. At 50~V, however, the calibration factor for the K-shell peak is of order of 10\% higher than that of the L-shell peak, which is attributed to saturation effects. Therefore, the energy scale is linearized before a final calibration factor is deduced (see text).
    }
    \label{fig:CalibrationFactors_det1}
\end{figure}

Figure~\ref{fig:CalibrationFactors_det1} summarizes the outcome of these fits for detector 1 (0~V and 50~V). For clarity, instead of the peak position itself, the figure shows the nominal interaction energy divided by the fitted peak position (which can be interpreted as a calibration factor), together with the uncertainties. At 0~V, these calibration factors are consistent with each other (which is also true for detector 3), which means that the peak positions are consistent with a linear calibration. The calibration at 0~V is then achieved by fitting a straight line (forced through the origin) to the calibration points.

However, at 50~V bias, for both detectors the calibration factors deduced from the K-shell peaks are significantly higher than the ones from the L-shell peaks, indicating a non-linearity on the order of 10\% at the K-shell energy. This is attributed to saturation effects, which lead to a distortion of the pulse shape and consequently to an underestimation of the interaction energy when using an energy-independent template in the OF algorithm. A simple pulse integral is linear in energy up to much higher energies as demonstrated below (Sec.~\ref{sec:high_calib}), but exhibits a worse resolution than the OF amplitude. To produce a linear energy scale, while taking advantage of the better resolution of the OF, we fit a logarithmic function of the form
\begin{equation}
\label{eq:linearization_rel_Eq}
    \text{A}_\text{OF} = a \log(b\cdot \text{A}_{\text{Int}} + 1)
\end{equation}
to the OF-versus-integral distribution, where A$_\text{OF}$ is the amplitude given by the OF, A$_\text{Int}$ is the pulse integral (divided by the time bin width) and $a$ and $b$ are fit parameters. We then invert that function to map the measured OF amplitudes onto the integral scale, producing an energy estimator that is linear up to at least the K-shell peak energy while maintaining the good resolution provided by the OF. The final calibration is then again achieved by fitting a straight line that is forced through the origin to the calibration data points.

\begin{figure}[pos=h]
    \centering
    \includegraphics[width=0.9\linewidth]{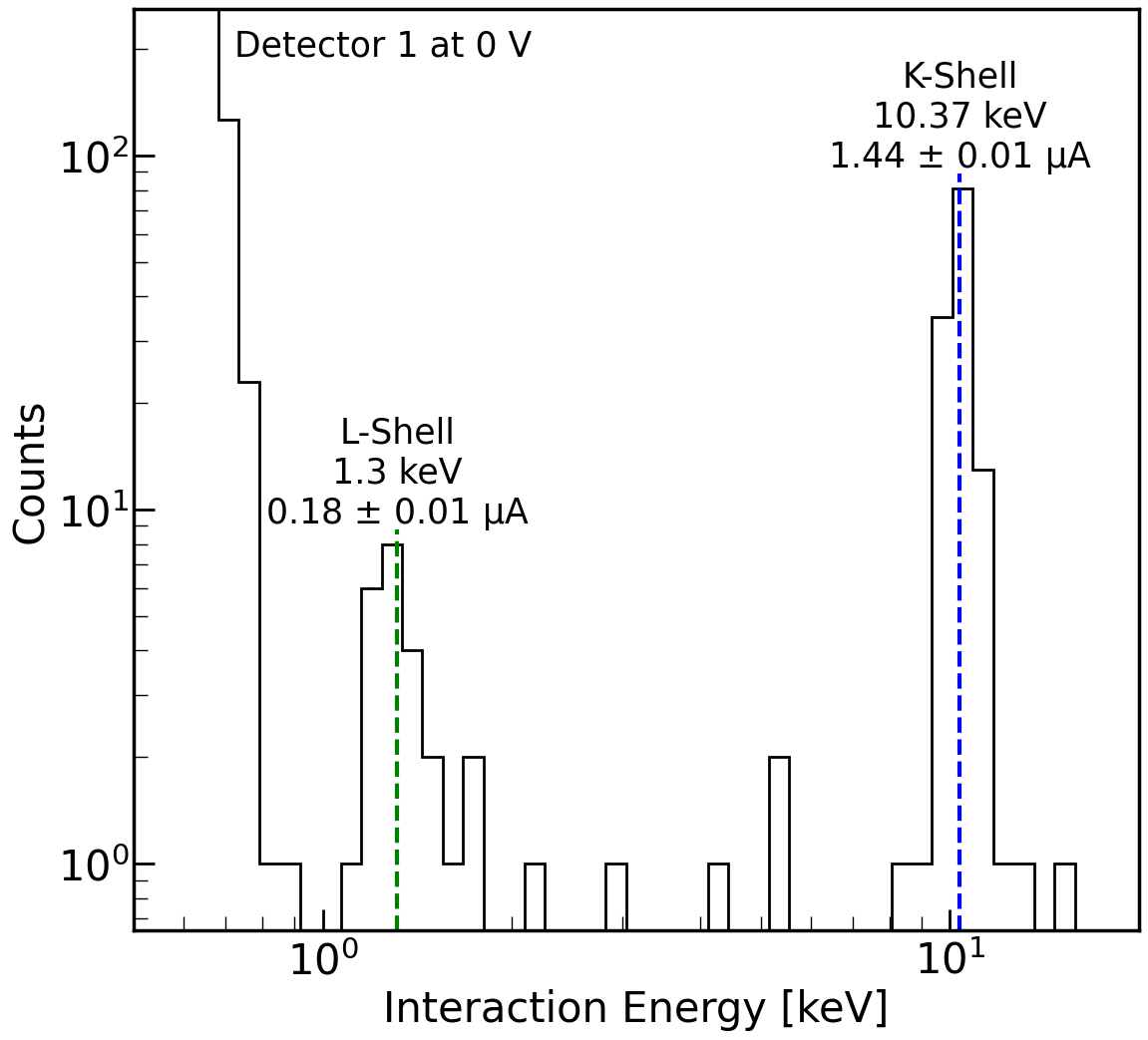}
    \includegraphics[width=0.9\linewidth]{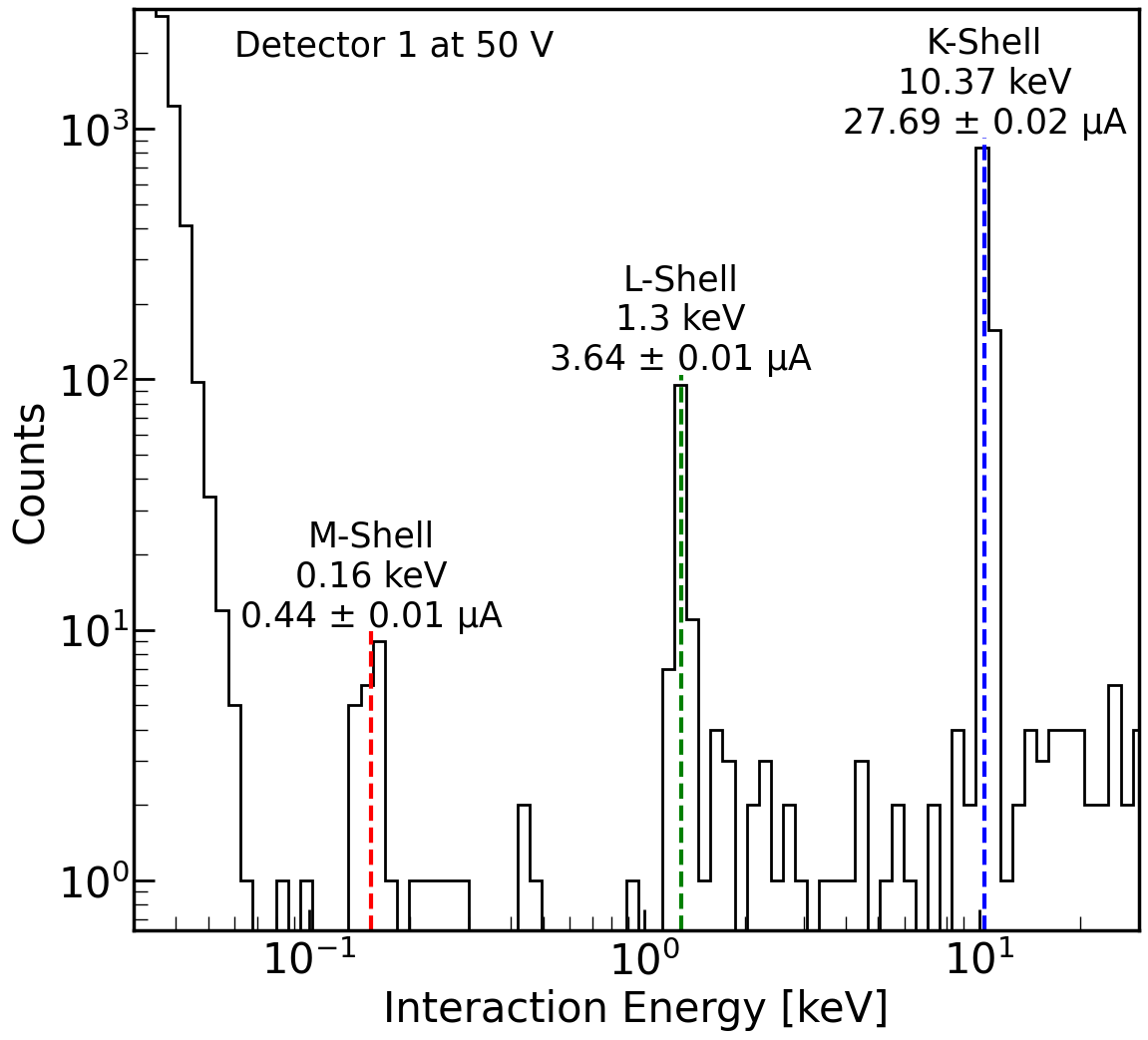}
    \caption{Energy spectra with $^{71}$Ge EC peaks for detector 1 at 0~V (top) and 50~V bias (bottom) after applying all basic and data quality selection criteria, as well as the low radius event selection (50~V data only). The nominal peak energies as well as the peak positions in the original (uncalibrated, non-linearized) spectra, as determined by our fit procedure, are indicated in the plots. 
    }
    \label{fig:OF_spec_lowE}
\end{figure}

Figure~\ref{fig:OF_spec_lowE} shows the calibrated OF spectra for detector~1 at 0~V bias (top) and 50~V bias (bottom). Table~\ref{tab:calibFactors} lists the relevant calibration and fit constants to convert the measured OF amplitudes to interaction energies.

\begin{table}[]
    \centering
    \caption{Calibration factors for the OF-based energy scale at 0~V ($\mathcal{C}_0$), and linearization constants $a$ and $b$ (see Eq.~\eqref{eq:linearization_rel_Eq}) and calibration factors for the linearized OF-based energy scale at 50~V ($\mathcal{C}_{50}$). The inverse of Eq.~\eqref{eq:linearization_rel_Eq} converts the OF scale into integral units, explaining the different magnitude of $\mathcal{C}_{50}$ when compared to what would be expected based on the voltage and $\mathcal{C}_{0}$. The listed uncertainties result from the propagation of the various fit uncertainties.} 
    \label{tab:calibFactors}    \setlength{\tabcolsep}{3pt} \renewcommand{\arraystretch}{1.5}
    \begin{tabular}{ccccccc}
    \toprule
        & \ & \textbf{0~V} & \ & \multicolumn{3}{c}{\textbf{50~V}} \\ 
    \midrule
    \textbf{Det.} &&$\mathcal{C}_0$&& $a$&   $b$    &$\mathcal{C}_{50}$\\
     \textbf{\#} &&  [keV/$\mu$A] &&[mA]&[1/$\mu$A]&[keV/mA]\\
    \midrule
    \textbf{1}&& 7.22$\pm$0.03 && 0.188$\pm$0.020 & 19.8$\pm$2.3 & 21.0$\pm$0.1\\ 
    \textbf{3}&& 4.03$\pm$0.01 && 0.223$\pm$0.024 & 19.2$\pm$2.4 & 13.9$\pm$0.2\\ 
    \textbf{6}&& 7.47$\pm$0.01 && -- & -- & -- \\ 
    \bottomrule
    \end{tabular}
\end{table}

\subsection{High Voltage Performance}
\label{sec:hv-performace}
\vspace{0.5em}
The benefit of applying a voltage bias across the detectors depends crucially on the detectors' response to this bias. At SuperCDMS Soudan, when selecting detectors for the CDMSlite operation \cite{CDMSlite1,CDMSlite2,CDMSlite3}, we found that at times some of our detectors showed signs of voltage breakdown (voltage-induced current through the Ge crystals) at voltages below 30~V. We also observed that the baseline noise may increase as a function of voltage. Finally, when operating Si HVeV detectors \cite{HVeV}, we found the signal amplification via the NTL effect to be larger than expected \cite{HVeV_Design_2021, HVeV_ComptonSteps}. All of these phenomena may have a critical impact on the dark matter sensitivity of our new detectors. Therefore, we obtained a series of data sets, operating all three Ge detectors discussed here under voltage biases ranging from 0~V to 90~V (referred to as voltage-scan data), to better understand their performance under bias. 

In the remainder of this section, we will discuss our findings regarding the NTL amplification and general observations regarding operations under voltage bias. The impact of voltage bias on the baseline resolution will be discussed in Sec.~\ref{sec:resolution}.

\begin{figure}[pos=h]
    \centering
    \includegraphics[width=\linewidth]{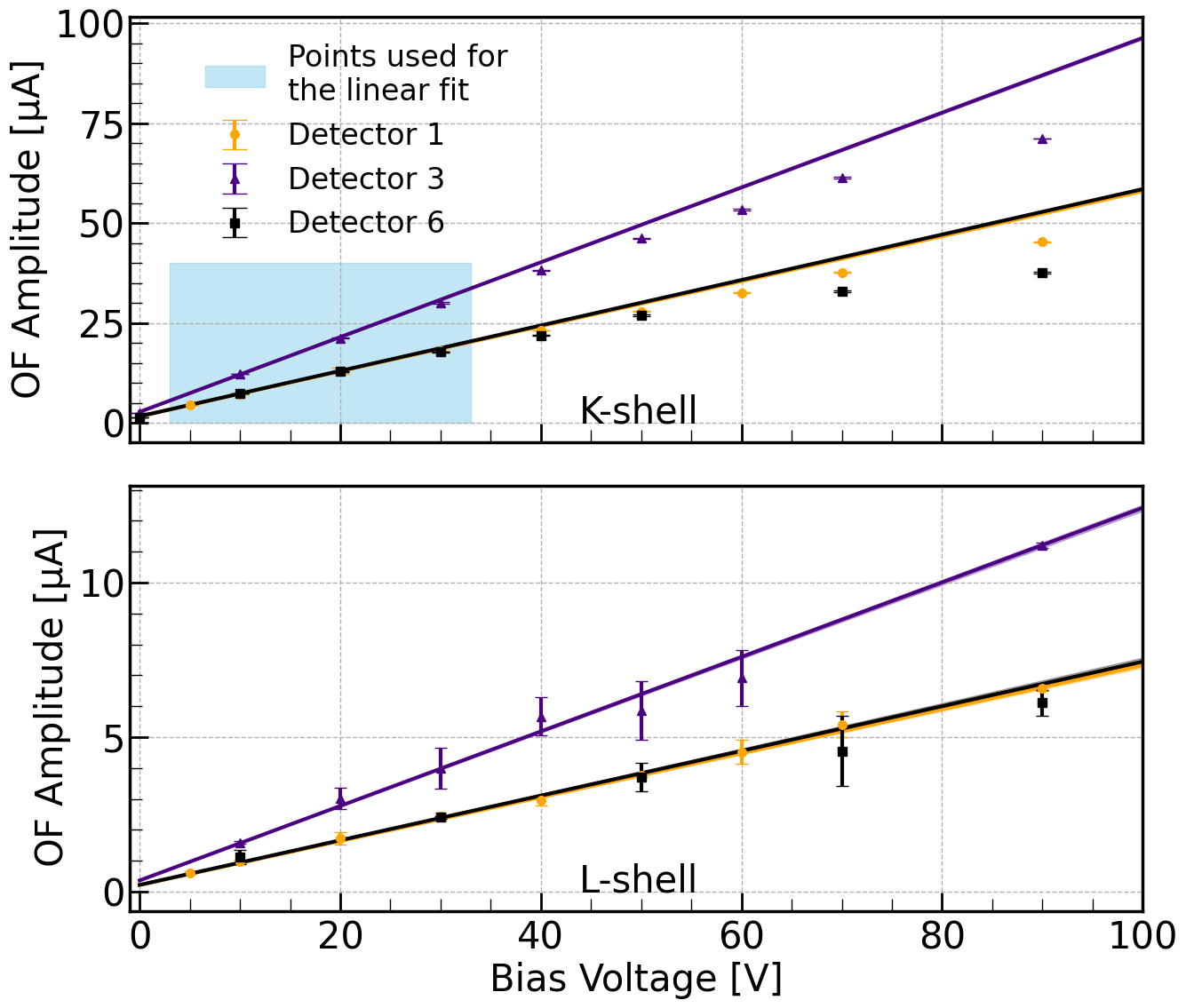}
    \caption{$^{71}$Ge EC peak position versus voltage, separately for K-shell (top) and L-shell peaks (bottom). For each detector, a line following Eq.~\ref{eq:NTL-amp} is fit to the data points acquired under non-zero bias voltages, where in case of the K-shell peaks, the fit range is limited to $\leq$30~V (light blue box) because the data points at higher voltage show a clear sub-linear trend which is attributed to saturation effects. The fit line for the detector 1 L-shell peak position (orange) is very similar to that for detector 6 and is therefore partially hidden behind the black line. Error bars represent the fit uncertainty of the peak position.
    }
    \label{fig:HV_scan_AvsV}
\end{figure}

According to Eq.~\eqref{eq:NTL-amp}, the total phonon energy (and with it the measured peak position) is expected to increase linearly with the applied voltage, with a slope proportional to the primary interaction energy E$_{\rm i}$. Figure~\ref{fig:HV_scan_AvsV} shows the peak position as a function of voltage, separately for the K-shell and L-shell peaks. The fit lines obey Eq.~\eqref{eq:NTL-amp}, though the equation uses calibrated and the plot uncalibrated energy estimators which explains the difference in slopes between the detectors. The L-shell data are consistent with expectations. The K-shell data points show a clear non-linearity at higher voltages which is attributed to saturation effects (note that the linearization procedure described in Sec.~\ref{sec:low_calib} has not been (and cannot be) applied here because at higher voltages, the integral-based energy estimator also becomes nonlinear as discussed in Sec.~\ref{sec:high_calib}). For the fit, only data points acquired under non-zero voltage bias of up to 30~V have been used.

\begin{figure}[pos=h]
    \centering
    \includegraphics[width=\linewidth]{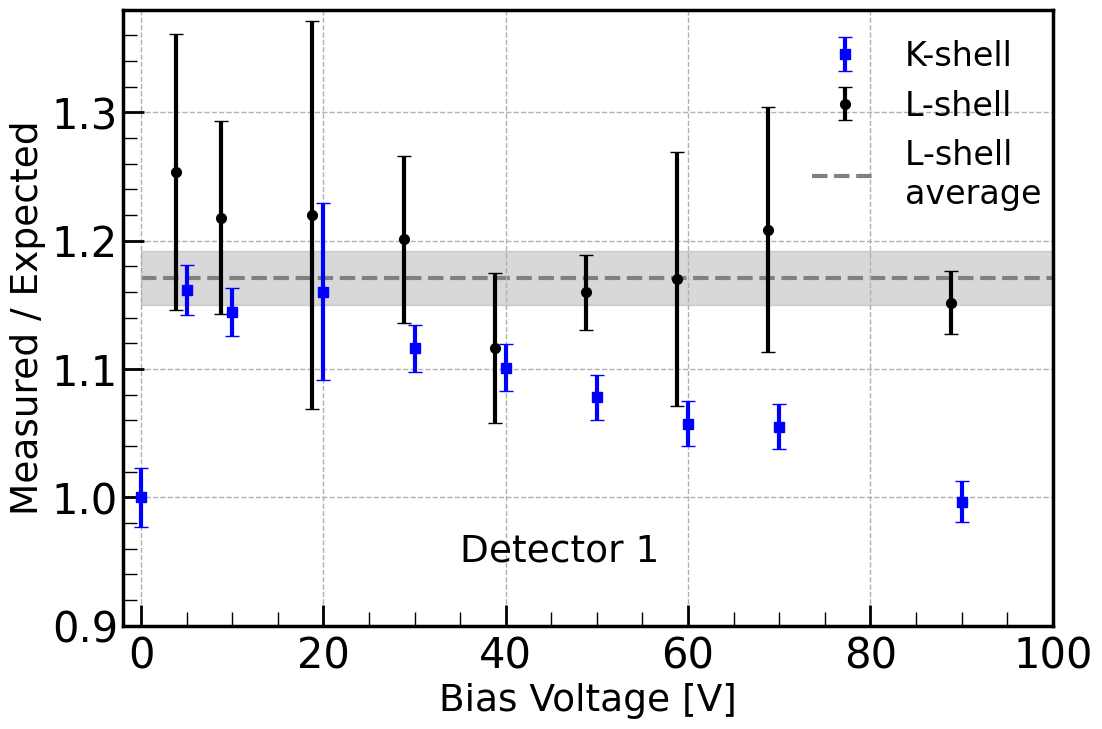}
    \caption{Measured $^{71}$Ge EC peak positions (calibrated using the 0~V K-peak position) divided by the expected peak position (according to Eq.~\eqref{eq:NTL-amp}) versus voltage. Data points for the L-shell (black) are shifted slightly to the left relative to the applied voltage to avoid overlap with the K-shell data points (blue). For the expected behavior, the data points would show a voltage-independent ratio of 1. The L-shell data points are consistent with a voltage-independent ratio, but at a higher value (best fit of 1.17 is shown as dashed line; the gray band represents the 1\,$\sigma$ uncertainty). The K-shell data points are systematically below the L-shell data points and show a decreasing trend with voltage, suggesting that they are impacted by partial saturation even at the lowest applied voltage of 5~V. This plot shows data for detector 1 only; data for the other two detectors are similar. Error bars are a combination of the fit uncertainty on the peak position and the 0~V calibration uncertainty.
    }
    \label{fig:HV_scan_ratio}
\end{figure}

Figure \ref{fig:HV_scan_ratio} shows the peak positions (K-shell and L-shell) in detector 1, first converted to energy using the 0~V calibration and then divided by the total expected phonon energy according to Eq.~\eqref{eq:NTL-amp}, as a function of the applied voltage. Data following this equation would show a voltage independent ratio of 1. The L-shell data points are consistent with being voltage-independent, but at a ratio larger than 1. This demonstrates an over-amplification similar to what has been observed previously in the Si HVeV detectors \cite{HVeV_Design_2021, HVeV_ComptonSteps}. The K-shell data points show a clear voltage dependence, which is attributed to saturation effects, as mentioned above; hence, we only draw conclusions from the L-shell data. Table~\ref{tab:overampl} lists the over-amplification factors determined from the L-shell data for each detector, as well as their weighted mean.  

\begin{table}
    \centering
    \caption{Over-amplification observed in the three detectors, determined by the L-shell peak position. The listed uncertainties result from the propagation of the involved fit uncertainties.}
    \label{tab:overampl}
    \setlength{\tabcolsep}{3pt}
    \begin{tabular}{lcccc}
        \toprule
        \textbf{Detector}&\textbf{1}&\textbf{3}&\textbf{6}&\textbf{Mean}\\
        \midrule
         Over-ampl. (\%) & 17.1$\pm$2.1
                                 & 13.9$\pm$3.4
                                            & 17.2$\pm$2.1&
                                                   16.6$\pm$1.8\\
        \bottomrule
    \end{tabular}
\end{table}

This is the first time this apparent over-amplification has been demonstrated in Ge detectors; it is also the first time that it could be shown that there is no significant voltage dependence over a wide range of voltage biases. This may be an indication of some signal-loss mechanism at 0~V and could potentially be related to charge carrier trapping, although further studies are required to draw firm conclusions (see also \cite{HVeV_SiO2} for related observations).

In some of our data sets taken under voltage bias, there is indication that the pre-biasing was not sufficiently high or long, leading to a residual elevated rate of low energy events that decreases over time and potentially has an impact on the baseline resolution assessment.

Detector 6 shows instabilities when operated at a bias of 30~V or higher. After an initial stable period on the order of a few hours, the rate of low energy events starts to increase substantially. The majority of these excess events are removed by the data quality selection criteria, and of the surviving excess events, a substantial fraction is usually removed when selecting low radius events. However, some of the excess events survive and, at times, the excess event rate strongly dominates the overall trigger rate. This observation is consistent with localized breakdown-like behavior caused by a high accumulation of space charge near the electrodes. This hypothesis is supported by an observation discussed below (see Sec.~\ref{sec:resolution}), which indicates that this detector may see a substantial rate of ionizing below-threshold events (much more than the other detectors). Such events might, e.g., be caused by IR photons, and they could generate such a space charge. With IR being a possible explanation, this observation is not necessarily predictive for SuperCDMS since the IR conditions differ from those in CUTE.

\subsection{Resolution}
\label{sec:resolution}
\vspace{0.5em}
During the measurement campaign, we discovered that the state and configuration of the DCRCs have a significant impact on the observed resolution. This is caused by high-frequency noise from the on-board clocks, which injects parasitic power into the sensors. We assessed the performance when operating multiple detectors simultaneously and when only operating a single detector at a time and found that single detector operation generally led to lower noise conditions and in turn a better energy resolution. However, we also found that certain readout channels on the DCRCs were particularly sensitive to external noise pickup, adding additional excess noise, and thus worsening the resolution.

At the beginning of each data set, we collect several hundred randomly triggered traces to provide the necessary noise information for the OF algorithm. These traces are also used to assess the baseline resolution. As discussed in Sec.~\ref{sec:reconstruction}, we use the zero-delay OF algorithm for this. The baseline resolution is determined as the standard deviation of a Gaussian function fitted to the distribution of the reconstructed amplitudes of these noise traces. This resolution in units of current is then converted to energy units using the calibration for the given data set (see Tab.~\ref{tab:calibFactors}). The best measured resolutions (in current units) for each detector were found in 0~V data sets during single detector operations; their energy equivalents are listed in Tab.~\ref{tab:baseline}. 

\begin{table}[]
    \centering
    \caption{Best measured baseline resolution ($1\,\sigma$) for each detector (single detector operation at 0~V). }
    \label{tab:baseline}   
    \begin{tabular}{rccc}
        \toprule
        \textbf{Detector}& \textbf{1} & \textbf{3} & \textbf{6}\\ \midrule 
         Resolution [eV] & 89.2 $\pm$ 0.7  & 80.2 $\pm$ 1.0  & 167 $\pm$ 14 \\
        \bottomrule 
    \end{tabular}
\end{table}

\begin{figure}[pos=h]
    \centering
    \includegraphics[width=1\linewidth]{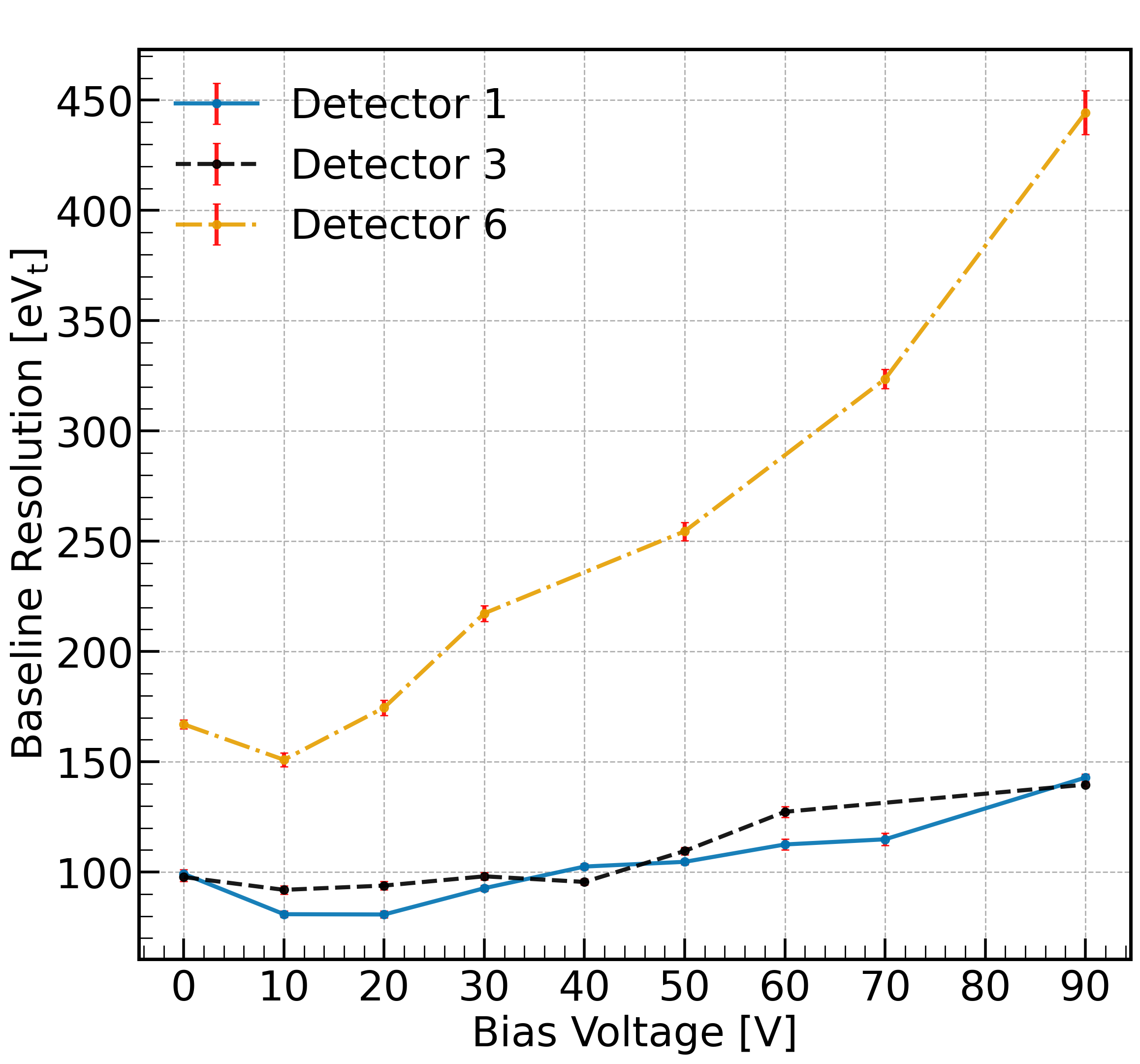}
    \caption{Baseline resolution ($1\,\sigma$), in total phonon energy units (eV$_\text{t}$), assuming standard NTL amplification as per Eq.~\eqref{eq:NTL-amp}, as a function of bias voltage. The error bars only account for the fit uncertainty in the respective energy spectra. Detectors 1 and 3 were operated simultaneously (worsening the resolution compared to the values in Tab.~\ref{tab:resolutionmodel}) while detector 6 was operated separately. Overall, we observe an increase in baseline resolution with voltage which is modest in detectors 1 and 3 and more significant in detector 6. This is an indication of a noise contribution from ionizing sub-threshold events (as could e.g.\ be caused by IR photons). The decrease in resolution between 0~V and 10~V in detectors 3 and 6 is a consequence of the over-amplification discussed in Sec.~\ref{sec:hv-performace}; in detector 1, the resolution at 10~V is lower than the 0~V resolution even before calibration; the origin here are systematic fluctuations in resolution measurements between individual data sets.
    }
    \label{fig:bl_bs_HV}
\end{figure}

\begin{figure}[pos=h]
    \centering
    \includegraphics[width=1\linewidth]{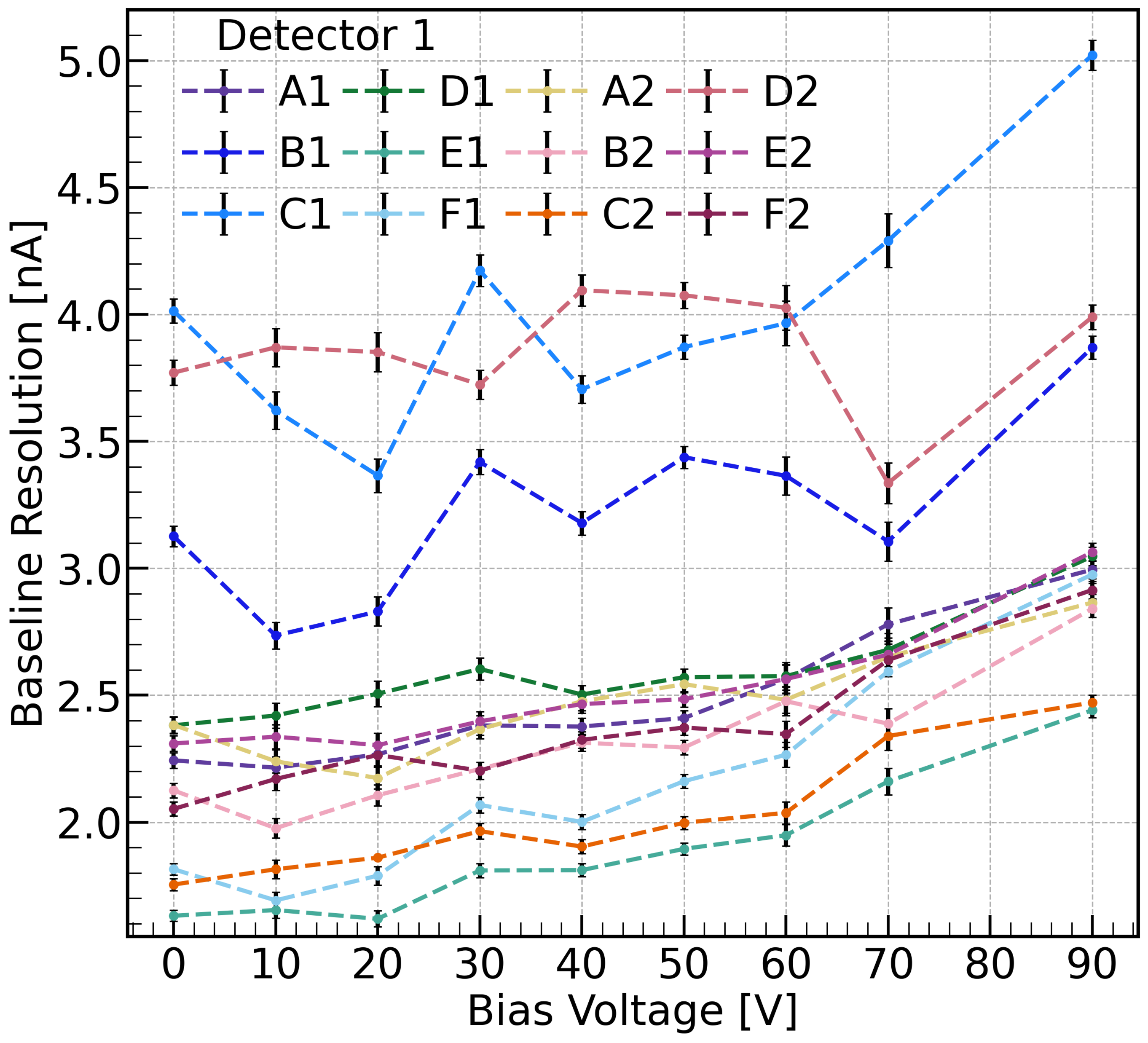}
    \caption{Baseline resolution ($1\,\sigma$) of individual channels in detector 1 (uncalibrated) versus\ bias voltage. Fluctuations significantly beyond the fit uncertainties (error bars) indicate the presence of large uncontrolled systematic contributions to the baseline resolution, likely related to variable noise pickup by the DCRCs.
    }
    \label{fig:bl_bs_HV_chan}
\end{figure}

Figure~\ref{fig:bl_bs_HV} shows the $1\,\sigma$ baseline resolution in total phonon energy units, calculated assuming standard NTL amplification (see Eq.~\eqref{eq:NTL-amp}), as a function of bias voltage for all detectors. Detectors 1 and 3 show a modest worsening of the resolution with increasing bias voltage, while in detector 6 the effect is more pronounced. This observation may be explained by a noise contribution originating from the presence of ionizing sub-threshold interactions at different rates in the different detectors, since this type of noise undergoes NTL amplification and thus worsens the resolution as a function of bias voltage. However, the NTL gain at 90~V is more than thirty, so when referred to interaction energy, we still see a significant improvement in resolution in all cases; for detectors 1 and 3, the 1~$\sigma$ resolution for the interaction energy at 90~V is about 4~eV, translating into a detection threshold of about 20~eV to 30~eV. In all three detectors, we observe an improvement in resolution between 0~V and 10~V. In detectors 3 and 6, this is caused by the over-amplification discussed in Sec.~\ref{sec:hv-performace}. In detector 1, however, the resolution at 10~V improves even before calibration. This is caused by systematic fluctuations in the resolution that go significantly beyond the fit uncertainties. Figure~\ref{fig:bl_bs_HV_chan} shows the (uncalibrated) baseline resolution for individual channels of detector 1. In addition to the very large differences in the overall magnitude between individual channels, there are large fluctuations within some of the channels. The leading hypothesis for these fluctuations is variable noise pickup by the DCRCs, since intrinsic noise sources are either voltage-independent, or monotonically worsen with voltage.

In addition to the indication of variable noise pickup by the DCRCs, we observed that high-frequency (out-of-band) noise generated by the DCRCs themselves has a noticeable impact on the detector performance. To mitigate the problems caused by excess noise (worsening resolution when operating multiple detectors, parasitic power from out-of-band noise from the DCRCs, noise pickup), additional electronic filter options have been introduced for SuperCDMS SNOLAB. If the additional filtering works as anticipated, all individual channels of a given detector should perform similarly, and at least as well as the best observed channel in the measurements reported here. This would correspond to a $1\,\sigma$ baseline resolution of the order of 70~eV, 50~eV and 90~eV or better for detectors 1, 3 and 6, respectively. This is getting close to the performance projected in Ref.\ \cite{SuperCDMS:Snowmass2021}; a 50~eV resolution would be sufficient to achieve the science goals that were set for these detectors, while a baseline resolution of 70~eV or 90~eV would restrict the accessible mass range at the lower end but would not prevent us from reaching world-leading sensitivity. There may be additional contributions to excess noise or parasitic power that were not identified. If the additional filtering or differences in experimental conditions between CUTE and SuperCDMS reduce such contributions, the resolution might improve even further. 

The total energy resolution $\sigma_t$ as function of energy can be modeled as

\begin{equation}
    \label{eq:resolutionmodel}
    \begin{split}
        \sigma_t = \sqrt{\sigma_{bl}^2 + \sigma_{pt}^2 + \sigma_{\text{F}'}^2}
        \\
                = \sqrt{\sigma_{bl}^2 + (A\text{E}_\text{i})^2 + \epsilon{\text{E}_\text{i} \text{F}'}}
    \end{split}
\end{equation}

\noindent where $\sigma_{bl}$ is the baseline resolution; $\sigma_{pt} = A\text{E}_\text{i}$ with the free model parameter $A$ accounts for all effects that contribute linearly in energy, such as position or time dependence of the energy response; and $\sigma_{\text{F}'}=\sqrt{\epsilon\text{E}_\text{i}\text{F}'}$ with the effective Fano factor F$'$ accounts for statistical uncertainties in the number of charge carriers. F$'$ replaces the intrinsic Fano factor F, which governs the uncertainty $\sigma_\text{F}=\sqrt{\epsilon \rm{E_i F}}$ in the number of eh pairs originally produced, where F$'\ge$F. This is to acknowledge that loss processes such as recombination or charge trapping may reduce the number of eh pairs, and thus increase the related uncertainty (and the possibility that additional effects may contribute to the uncertainty with the same energy dependence).

\begin{figure}[pos=h]
    \centering
    \includegraphics[trim={0.2cm 0.5cm 3.1cm 3.5cm},clip,width=0.99\linewidth]{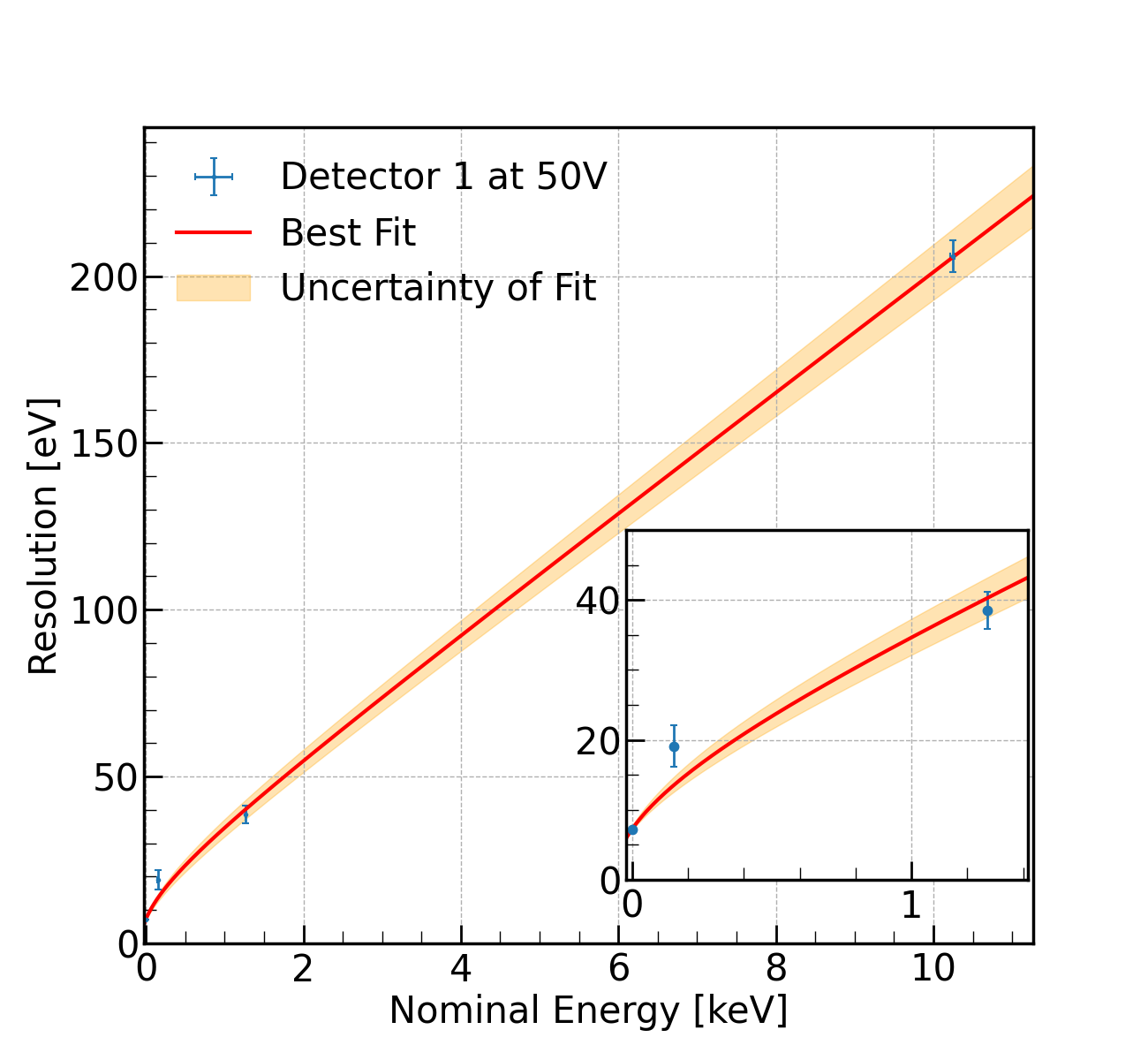}
    \caption{Resolution model fit (solid line) for detector 1 when operated under a bias of 50~V. The shaded band represents the $1\,\sigma$ fit uncertainty. The measured $1\,\sigma$ resolution values, the fit parameter $A$ and the effective Fano factor F$'$ for detectors 1 and 3 are listed in Tab.~\ref{tab:resolutionmodel}.
    }
    \label{fig:resolutionmodel}
\end{figure}

We only apply this model to our 50~V data sets, where, in addition to the baseline resolution, we have three data points from the measured $^{71}$Ge decay. Figure~\ref{fig:resolutionmodel} shows, as an example, the best-fit resolution model for detector~1. Table~\ref{tab:resolutionmodel} includes the measured resolutions and the fit parameters $A$ and F$'$ for detectors 1 and 3. 

For both detectors, the effective Fano factor is greater than Fano factors for Ge reported in literature (0.13 and lower, \cite{FanoFac_Bilger, FanoFac-refval, FanoFac_Sher}), which is expected since we operate our detectors at a relatively low electric field where loss processes are more prominent (an increase in measured Fano factors with decreasing electric field has been reported, e.g.\ in Ref.~\cite{FanoFac_Bilger}). For detector 1, the value of F$'$ is consistent with the effective Fano factor of 0.21-0.29 reported by CDMSlite  \cite{CDMSlite_R2long,CDMSlite3}, while that for detector 3 is approximately two standard deviations higher. 

\begin{table}[t]
    \centering
    \caption{Measured $1\,\sigma$ resolution and resolution model parameters for the 50~V data sets collected for detectors 1 and 3. The effective Fano factor for detector 1 is consistent with the value reported by CDMSlite~\cite{CDMSlite_R2long} while that for detector 3 is approximately two standard deviations away. The non-zero voltage bias and operating multiple detectors simultaneously contribute to worsening the baseline resolutions compared to the values reported in Tab.~\ref{tab:baseline}.
    }
    \label{tab:resolutionmodel}
    \resizebox{\columnwidth}{!}{
    \begin{tabular}{llcc}
         \toprule
         \textbf{Detector}&    &\textbf{1}        &\textbf{3}       \\ 
         \midrule
                   & K-shell & 206 $\pm$ 5\ \ \ & 232 $\pm$ 7\ \ \ \\
         Resolution& L-shell &   39  $\pm$ 3\ \ &   50 $\pm$ 4\ \ \\
         ($1\,\sigma$) \text{[eV]}
                   & M-shell &   19  $\pm$ 3\ \ &   19 $\pm$ 4\ \ \\
                   & Baseline&\ 7.22 $\pm$ 0.02 &\ 7.5 $\pm$ 0.1 \\
         \midrule
         Model      &$A$ ($\times 10^{-2}$) 
                           & 1.79 $\pm$ 0.08 & 1.9  $\pm$ 0.1 \\
         Parameters & \rm{F}$'$ & 0.28 $\pm$ 0.06 & 0.47 $\pm$ 0.10\\
        \bottomrule
    \end{tabular}
    }
\end{table}

The measured energy resolutions for the 0 V calibrations for all three detectors are listed in Tab.~\ref{tab:0V_resolution}

\begin{table}[t]
    \centering
    \caption{Measured $1\,\sigma$ resolution for the 0~V data sets collected for detectors 1, 3 and 6. The baseline resolutions here are worse than the ones reported in Tab.~\ref{tab:baseline} because here, multiple detectors were operated simultaneously.
    }
    \label{tab:0V_resolution}
    \resizebox{\columnwidth}{!}{
    \begin{tabular}{llccc}
         \toprule
         \textbf{Detector}& &\textbf{1} &\textbf{3} &\textbf{6} \\ 
         \midrule
         Resolution
            & K-shell &475 $\pm$ 30 &766 $\pm$ 41&533 $\pm$ 13\\
         ($1\,\sigma$) \text{[eV]}
            & L-shell &180 $\pm$ 30 &273 $\pm$ 46& \hspace{0.6em}-- \\
            & Baseline&125 $\pm$ 2\hspace*{0.4em}&128 $\pm$ 1\hspace*{0.4em}&212 $\pm$ 1\hspace*{0.4em}\\
        \bottomrule
    \end{tabular}
    }
\end{table}

\subsection{High Energy Calibration}\label{sec:high_calib}
\vspace{0.5em}
The low mass dark matter particles that SuperCDMS is focusing on would only deposit energies up to a few keV. Therefore, the HV detectors have phonon sensors optimized for low energy, with saturation effects starting to appear at energies as low as a few tens of keV of total phonon energy. Our initial assumption was that these detectors were not suitable for the detection of energies considerably above the saturation onset. However, we found that a simple trace integral provides an energy estimator that remains sensitive up to much higher energies than our standard OF-based energy estimator (though it is worse in resolution than the OF at low energy).

The high energy part of the spectrum that becomes accessible by this method can help us understand the radioactive environment the detectors are exposed to. This in turn makes it possible to constrain the background contribution in the energy range that is relevant to the dark matter search and thus improve the dark matter sensitivity.

Data were collected while the payload was exposed to a $^{133}$Ba gamma source. The source strength was too high for the source to be placed near the detectors; instead, it had to be retracted so that the line of sight between the source and the detectors was partially blocked by lead shielding located inside the cryostat, and in some cases by other detectors. As a consequence, the illumination is very inhomogeneous, with a majority of Ba-induced events occurring on one side of the detector near the outer edge. Detector~6 is almost completely blocked by the internal lead shielding and receives only indirect hits, while the geometry is most favorable for detector~1, which is the detector we discuss in this section. We collected Ba data with this detector at 0~V and 50~V bias. Due to the inhomogeneous event distribution discussed above, we do not apply the low radius event selection described in Sec.~\ref{sec:cuts} and accept the modest contribution of events with reduced NTL gain.

When using the relative calibration factors for the individual channels deduced from low energy interactions (see Sec.~\ref{sec:rel_cal}), the expected features in the Ba spectrum are visible but clearly compromised. Removing the relative calibration factors (by setting them to 1) markedly improves the spectrum. There may be room for further improvement by optimizing the relative calibration factors for saturated events, but for this first proof-of-principle study, we decided to simply remove the relative calibration factors.

The most prominent peak in the $^{133}$Ba spectrum is expected to be from 356~keV gamma rays. After applying all basic and data quality selection criteria, we calibrated the 0~V Ba spectrum by fitting a truncated Gaussian function to a narrow range around that peak and then rescaled the whole spectrum so the peak appears at 356~keV. We then fitted the other two clearly visible peaks in the thus calibrated spectrum in the same way and found their positions to agree within uncertainties with the nominal gamma energies: (382.3$\pm$1.7)~keV for an expected energy of 383.9~keV and (302.1$\pm$1.1)~keV for an expected energy of 302.9~keV (uncertainties are statistical only). This simple fitting method does not account for any non-flat background underneath the peaks and the impact that this may have on the extracted peak position. For a rough assessment of the systematic uncertainty this may introduce, we performed an alternative fit to the complete spectrum above $\sim$290~keV, with an ad-hoc model composed of three Gaussians, and a reverse sigmoid function to account for the background underneath the peaks. Although this model is not physically motivated, it renders an overall reasonable fit. The peak positions extracted by this method deviate by up to about 1\% from the values reported above. This indicates that the extracted peak positions are relatively robust, and -- given these positions -- that any non-linearities in this spectrum are likely at most at the percent level.

\begin{figure}[pos=h]
    \centering
    \includegraphics[width=1.0\linewidth]{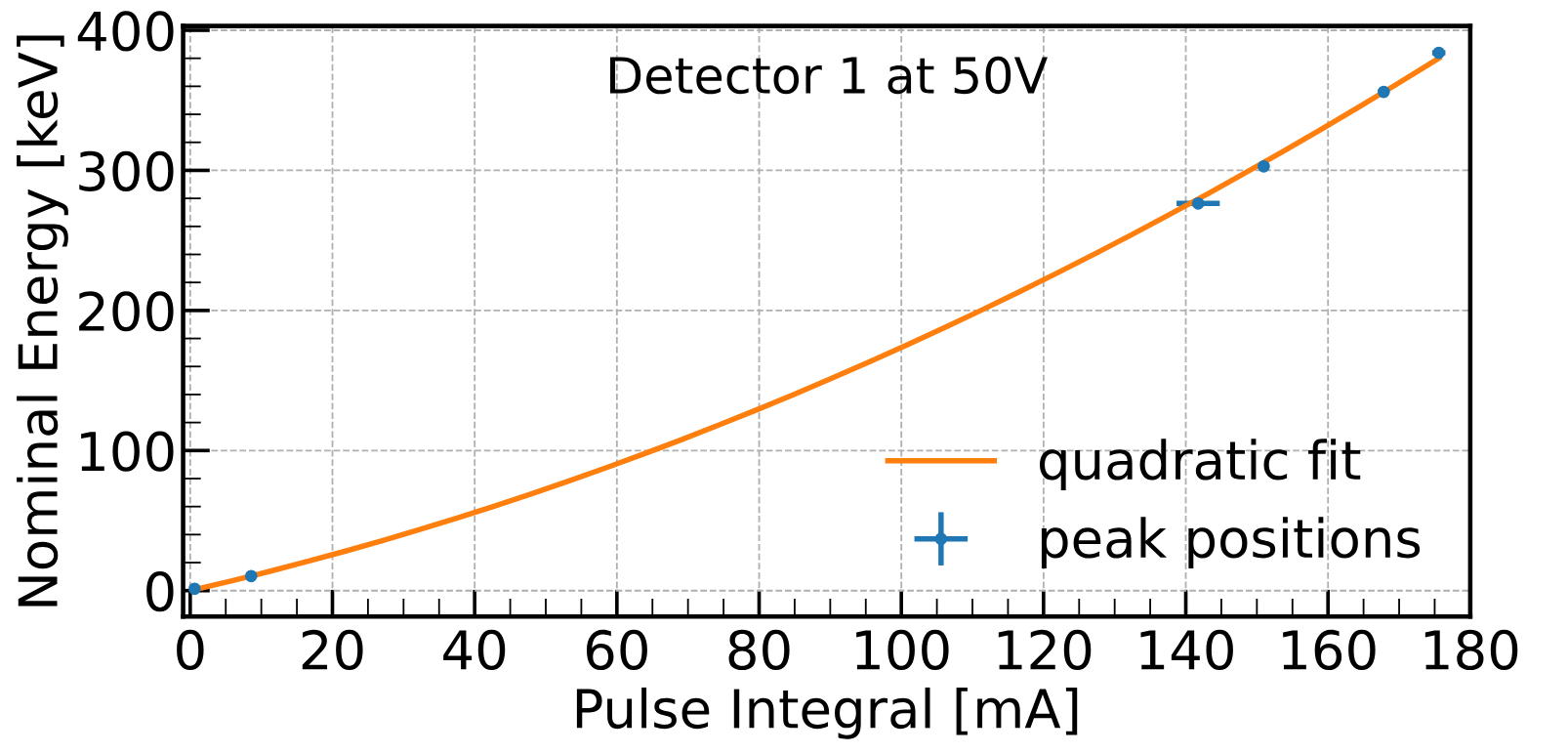}
    \caption{Calibration function for the 50~V data set collected with detector 1 after neutron activation and while being exposed to the $^{133}$Ba source. The fit function is a second order polynomial.
    }
    \label{fig:HV_calibrationFn}
\end{figure}

At a bias of 50~V across the detector, the energy response is clearly not linear anymore. Fortunately, in this case we have more data points available to determine a calibration function. At low energy, we see a peak in the integral spectrum corresponding to the K-shell capture events from $^{71}$Ge. When using the standard OF energy estimator, we can also identify the L-shell capture events. Thus, we have two low energy data points (the L-shell capture events are selected in the OF spectrum, and their integral distribution is then fit with a Gaussian). In addition, in the 50~V integral spectrum, the 276~keV peak from the Ba source is also visible and can be fitted. Figure~\ref{fig:HV_calibrationFn} shows the quadratic calibration function that maps the fitted peak positions to the corresponding energies. 

We estimate the systematic effects from non-flat backgrounds in a similar way as was done for the 0~V spectrum, while here we also consider the effects of events with reduced NTL gain on the spectrum. We fit the spectrum above about 250~keV using again a reverse sigmoid function for the background; the peaks are fit by single-sided crystal-ball functions. This allows the model to account for low energy tails to the peaks caused by events with reduced NTL gain. Such events appear near the edge of the detector due to the distorted electric field. The deviations of the extracted peak position from those found by the simple fit procedure described before are well below 1\%, except for the poorly constrained peak at 276.4~keV where a deviation of up to 2\% is seen.

\begin{figure}[pos=h]
    \centering
    \includegraphics[width=1.0 \linewidth]{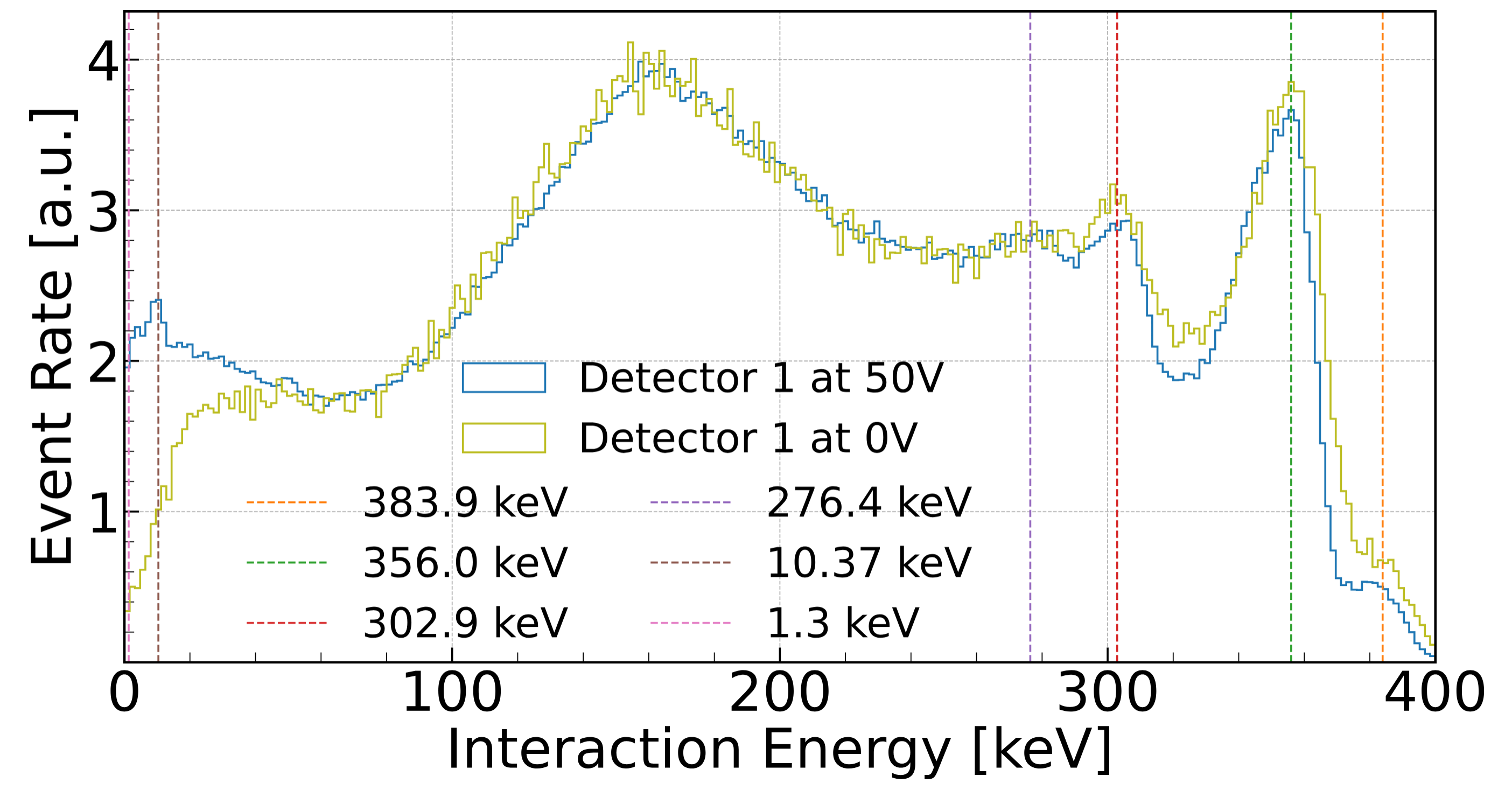}
    \caption{Calibrated energy spectra up to 400~keV interaction energy for detector~1, operated at 0~V (green) and 50~V bias (blue). Dashed lines indicate the energies of the $^{133}$Ba gamma emission lines as well as the K-shell and L-shell capture lines from the $^{71}$Ge decay. The backscatter peak at $\sim$160~keV results from gammas scattering in the surrounding material before entering the detector; its position and shape depends on the geometry of the setup.
    }
    \label{fig:HE_Ba_spectrum_0_50V}
\end{figure}

Figure~\ref{fig:HE_Ba_spectrum_0_50V} shows the two calibrated Ba spectra, accumulated with 0~V and 50~V bias across the detector, respectively. The resolution at 356~keV (1 $\sigma$) is better than 3\%. The deviation between the two spectra at low energy (below $\sim$50~keV) is caused by differences in trigger conditions, affecting events that are shared between different detectors.

\section{Conclusion}
\label{sec:conclusion}

The performance of three of the new SuperCDMS Ge HV detectors has been studied under low-background conditions underground at the CUTE facility, and calibrations at both low and high energy have been performed. 

Low energy calibration was achieved, exploiting the peaks in the energy spectrum generated by the electron capture decay of $^{71}$Ge, which had been produced in situ through neutron activation of the detectors with neutrons from a $^{252}$Cf source. Based on this calibration, the best measured $1\,\sigma$ baseline resolutions for detectors 1, 3 and 6 were found to be 89~eV, 80~eV and 167~eV, respectively (Tab.~\ref{tab:baseline}). This performance was only achieved when operating detectors individually; when operating multiple detectors simultaneously, interference from the readout electronics noticeably worsened the measured resolution. However, even in single detector readout mode, we have strong evidence that the observed baseline resolution is dominated by excess electronic noise in some of the readout channels. Consequently, additional filtering options have been developed and implemented for SuperCDMS SNOLAB to reduce the noise and allow multiple detectors to be operated simultaneously without a noise penalty. An increase in baseline resolution when applying bias voltage suggests that ionizing sub-threshold events (as may, e.g., be caused by IR photons) also contribute to excess noise, in particular at high voltage, and much more so in detector 6 than in the other two. In detector 6, we also saw a hint of sensitivity to residual vibrations in the system. If the excess noise observed in these measurements can be mitigated, resolutions in the range of 50~eV or better may be reachable, which would be sufficient to achieve the science goal for these detectors.

The voltage dependence of the detector response was carefully examined. It has been confirmed that all three investigated detectors could be operated at up to 90~V (the highest voltage tested) when properly neutralized. Studying amplification as a function of the applied bias voltage revealed that while the behavior at non-zero bias voltages follows the expected linear trend, an over-amplification is observed relative to the expectation based on 0~V amplitudes, similar to previous findings in the Si HVeV detectors \cite{HVeV_Design_2021, HVeV_ComptonSteps}.

The design of the SuperCDMS HV detectors is optimized for low energy interactions, and the sensors show strong non-linear behavior already at relatively low energies. It was therefore a surprise when we discovered that, using the pulse integral as energy estimator, we can achieve a resolution of better than 3\% at 356 keV (corresponding to nearly 7~MeV of total phonon energy for the data set taken with 50~V bias), as we have shown with the high energy calibration performed using a $^{133}$Ba gamma source.

Overall, we conclude that adequate methods are at hand to calibrate the new SuperCDMS Ge HV detectors, not only at the low energy range relevant for dark matter interactions, but up to hundreds of keV of interaction energy, allowing for the direct measurement of high energy gamma ray backgrounds which in turn will let us deduce their contribution to the background rate at low energies. With adequate measures in place to mitigate the excess noise observed here, the detectors will be able to achieve the performance needed to reach the scientific goals that were set out for SuperCDMS SNOLAB.

\section*{Acknowledgments}

The SuperCDMS collaboration gratefully acknowledges SNOLAB and its staff for providing underground space and logistical and technical support. Funding and support were received from the National Science Foundation, the U.S. Department of Energy (DOE), Fermilab URA Visiting Scholar Grant No.\ 15-S-33, NSERC Canada, the Canada First Research Excellence Fund via the Arthur B.\ McDonald Institute (Canada), the Department of Atomic Energy of the Government of India (DAE), the J.\ C.\ Bose Fellowship grant of the Anusandhan National Research Foundation (ANRF, India), the DFG (Germany) under 420484612 and under Germany’s Excellence Strategy – EXC 2121 – 390833306, and the Marie-Curie program – Contract No.\ 101104484. SNOLAB operations are supported by the Canada Foundation for Innovation and the Province of Ontario, with underground access provided by Vale Canada Limited at the Creighton mine site. Fermilab is operated by Fermi Forward Discovery Group, LLC,  SLAC is operated by Stanford University, and PNNL is operated by the Battelle Memorial Institute for the U.S.\ Department of Energy under contracts DE-AC02-37407CH11359, DE-AC02-76SF00515, and DE-AC05-76RL01830, respectively. This research was enabled in part by support provided by Compute Ontario \href{https://www.computeontario.ca/}{(computeontario.ca)} and the Digital Research Alliance of Canada \href{https://alliancecan.ca/en}{(alliancecan.ca)}.



\bibliographystyle{apsrev4-2}

\bibliography{cas-refs.bib}



\end{document}